\documentclass[final,3p,times,authoryear]{elsarticle}
\usepackage[utf8]{inputenc}

\usepackage{hyperref}
\usepackage{breakurl}

\usepackage{color,ifthen}
\usepackage{xcolor}

\usepackage[shortlabels]{enumitem}

\usepackage{pdfpages}

\journal{Springer's book ``Human Factors in Privacy Research''}
\date{August 2023}

\begin{document}

\begin{frontmatter}

\title{Expert opinions on making GDPR usable%
}

\author[uio]{Johanna Johansen\fnref{labelThanksx}}
\ead{johanna@johansenresearch.info}
\address[uio]{Department of Computer Science, Norwegian University of Science and Technology.}
\fntext[labelThanks]{I would like to thank Professor Simone Fischer-H\"{u}bner for great guidance and help with the present research work.}

\begin{abstract}
We present the results of a study done in order to validate concepts and methods that have been introduced in \cite{johansen2020making}. 
We use as respondents in our interviews experts working across fields of relevance to these concepts, including law and data protection/privacy, certifications and standardization, and usability (as studied in the field of Human-Computer Interaction). 
We study the experts' opinions about four new concepts, namely: 
(i) a definition of Usable Privacy, 
(ii) 30 Usable Privacy Goals identified as excerpts from the GDPR (European General Data Protection Regulation),
(iii) a set of 25 corresponding Usable Privacy Criteria together with their multiple measurable sub-criteria,
and (iv) the Usable Privacy Cube model, which puts all these together with the EuroPriSe certification criteria, with the purpose of making explicit several aspects of certification processes such as orderings of criteria, interactions between these, different stakeholder perspectives, and context of use/processing.

The expert opinions are varied, example-rich, and forward-looking, which gives a impressive list of open problems where the above four concepts can work as a foundation for further developments.
We employed a critical qualitative research, using theory triangulation to analyze the data representing three groups of experts, categorized as `certifications', `law', and `usability', coming both from industry and academia. 
The results of our analysis show agreement among the experts about the need for evaluations and measuring of usability of privacy in order to allow for exercising data subjects' rights and to evaluate the degree to which data controllers comply with the data protection principles. However, the community still needs to find archetypal usability thresholds to guide both the businesses in what would be an acceptable level to attain and for the evaluators to assess the level of compliance against.

Regarding the above four concepts our results first show that, while agreeing with the Usable Privacy definition, the experts are more often interested in finding (and giving) instances of this definition, i.e., 
examples of what are the specific areas of privacy or data protection that usability is most relevant for, or what are the goals and characteristics of the users that have particular implications for privacy.
Similarly, the experts are happy with the list of Usable Privacy Goals, seeing these as instances of the Usable Privacy definition.
However, even if the Usable Privacy Criteria are seen as a good solution for starting to evaluate usability of privacy, these are not giving enough detail and method for a data controller to be able to understand how (what usability techniques and processes to use) to meet these criteria.
Finally, the validation of the Usable Privacy Cube shows that this model captures at an abstract level many of the principles that are present in the existing processes of privacy evaluations. However, these are present more in an implicit manner, and not as determining factors. Nevertheless, since these concepts are being familiar to the evaluators, and being already part of their practice, further work will be to make these weight more towards considering if an organization has achieved compliance or not. In addition, the movement towards achieving usability goals should also inspire a shift in perspectives from ``what is enough to be compliant'' towards ``what would be the level of data protection that I would like to achieve in order to have a better competitive advantage''.
\end{abstract}

\begin{keyword}
Usable privacy \sep  General Data Protection Regulation \sep Expert opinions \sep  Validation \sep Certification \sep Qualitative research

\end{keyword}

\end{frontmatter}

\section{Introduction}
Since electronic privacy is a complex concept, with implications not only for individuals, but also for society at large, we have argued in \cite{johansen2020making,johansen2020theTR} that a multidisciplinary and pragmatic approach to usable privacy is needed. Particularly, \cite{johansen2020making} addresses the problem that existing data protection regulations, specifically the European General Data Protection Regulation (GDPR), are rather vague when describing the extent to which data protection principles and data subjects rights should be implemented as to be considered beneficial for the user. Therefore, we work towards a methodology that would produce measurable evaluations of the usability with which privacy goals of data protection are reached. Having a scale showing \emph{how well} a product respects the privacy of its users, and \emph{how easy} it is for the user to understand the level of privacy protection that a product offers, works towards fulfilling the goal expressed in the Recital (100) of the GDPR, i.e., that of ``allowing data subjects to quickly assess the \emph{level} of data protection of relevant products and services''.

Working towards the above expressed goals, we contribute in \cite{johansen2020making,johansen2020theTR} with defining, addressing and evaluating usability aspects of data protection. 

First, \cite{johansen2020making} proposes a \textit{definition of Usable Privacy}%
\footnote{A short video-presentation of the definition of Usable Privacy, which was also used during our interviews conducted with one of the groups of participants in this study, can be found at the following link: \url{https://vimeo.com/569510999}.}
which extends and adapts the definition of usability from the ISO 9241-11:2018  \cite{ISO9241-11:2018} to privacy. The rest of that paper shows how to apply this definition to the GDPR.
\begin{quote}
\textit{Usable privacy} refers to the extent to which a product or a service protects the privacy of the users in an efficient, effective and satisfactory way by taking into consideration the particular characteristics of the users, goals, tasks, resources, and the technical, physical, social, cultural, and organizational environments in which the product/service is used.
\end{quote}

Second, a model called the Usable Privacy Cube (UP Cube)%
\footnote{A short video-presentation of the UP Cube model that was used during the interviews with one of the groups of participants in our study can be found at the link: \url{https://vimeo.com/571358474} .} 
model is proposed in \cite{johansen2020making} to guide the process of evaluation of usable privacy in certifications. The UP Cube model has three axes of variability; the two at the base contain the existing EuroPriSe (European certification body) criteria, reorganized in two categories, i.e.: 
\begin{enumerate}[(i)]
\item rights of the data subjects, and 
\item data protection principles.
\end{enumerate}
These also represent the two perspectives on privacy that one usually takes, i.e.: 
\begin{enumerate}[(i)]
\item that of the users of which private information is being collected (and which the regulations usually seek to protect), and
\item that of the industry/controllers developing products or services that collect and process private information (and which must conform with regulations such as GDPR and show compliance by going through certifications such as the EuroPriSe). 
\end{enumerate}
The third vertical axis is composed of Usable Privacy Criteria intended for measuring the usability level of privacy in a specific context of use. 
The UP Cube model comes with other concepts that are beneficial for a certification process, such as the idea of ordering and prioritization of the criteria, as well as the possibility to identify intersections between the axes. 

Third, \cite{johansen2020theTR} lists 30 Usable Privacy Goals (UP Goals)%
\footnote{The UP Goals are presented in short video that was used during the interviews in our current study, and can be found at the link: \url{https://vimeo.com/569510999} .}
extracted from the GDPR text.
One such goal is, e.g., found in the Article 12:
\begin{quote}
``\dots any information \dots and communication \dots relating to processing [to be provided] to the data subject in a \emph{concise}, \emph{transparent}, \emph{intelligible} and \emph{easily accessible form}, using \emph{clear and plain language,} \dots''
\end{quote} 
How concise, transparent, or intelligible the form of presentation is, can be determined by measurements of efficiency, effectivity, and satisfaction, in a respective context of use. 
The emphasized words are those that can be interpreted differently based on the context they are used in, and can result in objective and perceived measurements when evaluated using usability methods. 

Finally, based on UP Goals, \cite{johansen2020theTR} formulates a set of Usable Privacy Criteria (UP Criteria)%
\footnote{The UP Criteria are presented in short video that was used during the interviews in our current study, and can be found at the link: \url{https://vimeo.com/556133682} .} meant to produce measurable evaluations of usability of privacy that can be translated into scales to be used in certifications. 
The goal from above, for example, is associated a criteria that contains several specific sub-criteria worded so to produce measurements, e.g., the one below would be requiring to measure efficiency:
\begin{quote}
How much time/effort/financial and material resources does the data subject need to invest in order to access the information related to the processing of his/her personal data?                                                                                                                                                                                    \end{quote}

\subsection{Our Study}\label{subsec_studyQuestions}

We would like through the present study to understand how experts in fields relevant for the above topics are currently dealing with usability aspects of the data protection. Moreover, we are particularly interested whether the solutions proposed in \cite{johansen2020making,johansen2020theTR} are aligned with their needs and practices. Summed up, the concepts that we aim to validate are:
\begin{enumerate}[(i)]
\item the Usable Privacy Definition, 
\item the Usable Privacy Criteria,  
\item the Usable Privacy Goals, and
\item the Usable Privacy Cube model. 
\end{enumerate}

We have created three groups of participants, described in detail in Section~\ref{sec_participants}. Each of these groups are formed around a specific expertise and are dedicated to validate one/two of the above four concepts as follows: 
\begin{enumerate}[(A)]
\item the `usability group' is used to study/validate the Usable Privacy Definition and the Usable Privacy Criteria,
\item the `certifications group' is used for the Usable Privacy Cube model, and 
\item the `law group' for the Usable Privacy Goals. 
\end{enumerate}

We follow the classical way of structuring a qualitative research based study, by presenting how we collected and analysed the data, as well as information about our participants in  a Method section (\ref{sec_method}). We continue with presenting our results in Section~\ref{sec_results}, where we conclude with Discussions. In the Results section we have a sub-section for each of our main key themes. We conclude in Section~\ref{sec_conclusion}, where we also mention further work and limitations.

\section{Method}\label{sec_method}
To validate the four new concepts listed in Section~\ref{subsec_studyQuestions} we employ a critical qualitative research \cite{braun2013successful}, where we take an interrogative stance towards the meanings and experiences expressed in the data we collect through interviews. We are also interested in how the individual meanings reflect how usability is understood by the broader communities that the participant's expertise are representing. To achieve this we involve three different theoretical perspectives in a ``theory triangulation'' manner \cite{patton1999enhancing}: certifications, law, and usability (Human-Computer Interaction/Interaction Design/UX experts). 
Special in our case is that the participants are not brought to discuss the data, but to discuss the theories and concepts introduced in \cite{johansen2020making,johansen2020theTR}. Their meanings then represent the data that we analyze in the rest of this paper.

With this approach we seek to validate our knowledge within the scientific and practice community \cite{angen2000evaluating}, as represented by the experts we brought into the discussion. We aim at bringing forth a ``disciplinary matrix'' \cite{mishler1990validation} of assumptions, theories, and practices shared on the topic of usability of data protection within the community of specialists on certifications, on data protection, and on usable privacy and security. To validate our knowledge claims, we created through interviews, an environment where a slice of the research and practice community could present their perspectives on the subject \cite{kvale1994interviews}. The perspectives are then analyzed in order to identify conflicting or agreeing interpretations, as well as possibilities for future development of the knowledge we claim in \cite{johansen2020making,johansen2020theTR}. Based on the feedback from the experts we seek to confirm%
\footnote{Validation is defined, in the Oxford English Dictionary, as the process of confirmation.} 
whether our ideas are valuable and useful enough to study, and whether the approach we adopted to the research questions we address is the right one. We intend to find out whether our research is both relevant and beneficial to those concerned, such as the certifications experts. Intending to reach both an ethical and substantive validation of our work, we have recruited people who have had experience with the topic, establishing a cooperative approach between the researcher and the researched in a social constructivism manner, which in our case is the community working with certifications \cite{angen2000evaluating}.

\subsection{Collecting interview data}

We chose to use the method of conducting interviews for collecting our data because this method is best suited when needing, as in our case, to explore understandings, perceptions, and constructions (which for us are of topics related to usability in data protection). We expect to generate rich and detailed responses because the participants that we chose have a personal stake in these topics, since they are working with privacy certifications and standards, and they need to address in one way or another aspects that we talk about in the interviews. Since \cite{johansen2020making} uses methods and terminology from the field of Ergonomics of human-system interaction to evaluate usability of data protection, we also invited experts from this field, especially those that have been working at the interaction between usability and privacy, e.g., from the field known as usable privacy and security. 

We had meetings of 40 to 60 minutes duration with each of the interviewees. The form of data collection employed in this study was approved by Sikt – Norwegian Agency for Shared Services in Education and Research\footnote{\url{https://sikt.no/en/about-sikt} -- previously known as NSD -- Norwegian Centre for Research Data} and each participant was asked to sign a consent form prior to the interview. The interview was held online, using Zoom, and the conversation was audio recorded. The data was transcribed and anonymized and the recording deleted after transcription for maintaining the anonymity of our participants.

We designed our interviews to be semi-structured, having a list of questions to guide the conversation, while the participants were encouraged to talk freely on the main topics of the interview. The topics and questions were adapted to the different expertise the participants have, concerning different aspects of our paper. 
All three interview types had two main parts. 
First we want to learn about the participants' current understanding of (and their relation with) the usability aspects of the data protection legislation, without biasing them with our views. 
Second, after we introduce our research through a short video presentation, we then asked the participants to express their opinion directly in relation with what we have presented. Before the interview, we informed the participants only about our general topic of research, as we did not want to influence them with our opinions. We also wanted spontaneous, and not preconceived responses, so to reflect ingrained knowledge of the respective fields and areas of practice the participants represent. 

We started the interview with a common topic for all three groups, where the participants presented their understanding and experience with usability in data protection (see Appendix~\ref{appendix_perspectiveExperienceOnUsability_q}). This was done with the intention to reveal the current understanding of usability in the respective domains, whether there are differences or overlaps between them, and also to see how their perspectives are (or could be) related to our definition of `usable privacy'. 
Afterwards we had specific topics for each of the groups:
\begin{itemize}
\item With the `certifications group' we discussed topics related to evaluating and measuring usability, as well as the Usable Privacy Cube model, because these participants have knowledge on which processes and methods are currently used in evaluation and certification processes. For the exact topics and questions see Appendix~\ref{appendix_evaluatingMeasuring} and \ref{appendix_UPCube}.

\item  The `usability group' addressed topics related to defining usable privacy 
and Usable Privacy Criteria, because this group is acquainted with the ISO 9241-11:2018 standard on usability that was used as a basis for the definition of usable privacy in \cite{johansen2020making}. Moreover, this group  knows well methods and processes of evaluating usability of digital products in general, as well as the process of formulating goals into criteria of evaluations. For the exact topics and questions see Appendix~\ref{appendix_UPDefinition} and \ref{appendix_UPCriteria}

\item The participants in the `law group', being well acquainted with the GDPR text, were asked to check if the list of the Usable Privacy Goals, appearing in \cite{johansen2020theTR}, is complete, and whether the goals were correctly chosen to represent usability aspects. For the exact topics and questions see Appendix~\ref{appendix_UPGoals}.
\end{itemize}

\subsection{The participants}\label{sec_participants}
The participants were sampled using convenience and snowball methods. The three groups, named `certifications group', `law group' and `usability group', 
were establish based on the expertise. However, most of the participants have a composite background, a mixture of computer science, law, and human factors. Common for all is that they are working (or doing research) on aspects related to electronic privacy and European data protection applied to IT services/products, thus all having knowledge of data and technology. 

Another criterion in our selection process was diversity among the participants, aiming to bring expertise both from industry and academia, as well as age and gender diversity.
Even though our initial recruiting of the participants tried to fulfill these criteria,
in the interviews 
we also asked the participants to specify their primary and secondary area of expertise, as well as their work experience (type of position held, type of organization, and number of years) relevant for data protection/privacy (see Appendix~\ref{appendix_demographics_q}).
One conclusion from the interviews, which strengthens one of our initial choices, is that all the participants in our study had some experience with usability related aspects from their work or at least a basic understanding of what usability is. This is apparent from the answers given to the first topic of our interview, which asked the participants to present their understanding and experience with usability. We attempted one interview with a participant that had expertise in data protection for IT, but was not acquainted with the concept of usability at all, which in the end forced us to remove this participant because most of the questions could not be answered.

The `certifications group' consists of six people working with standards, certifications, and data protection organizations. This is confirmed by their answers: 4 out of 6 have this as their main field of expertise  whereas the remaining two are working for Data Protection Authorities (DPAs). Moreover, all these participants have Law/data protection as part of their expertise (one as primary and 5 as secondary). The years of experience range from 6 to 32, and the gender is equally represented. 
The work experience ranges from leadership and research for DPAs, consulting, audit, or technical assessment for certification bodies and other governmental organizations, or board membership and other functions for standardization committees. We consider these backgrounds to represent well our target group.

One of the participants could have been placed, based on the main field of expertise, in the `usability group', but a person with a Usability/HCI/IxD/UX background combined with certification is rare and we wanted to have this participant contribute to the topics especially chosen for the `certifications group'. 
As detailed later, the `usability group' does not include experience with certifications.

The `usability group' contains seven people working with Usability (sometimes referred to by the name of the broader fields HCI/IxD/UX), being confirmed also by their answers, i.e., 6 out of 7 have this as their main field of expertise. Their secondary expertise was somewhat more diverse, including law/data protection, privacy and security, cybersecurity, contract design, design thinking, and Information Systems Development from an organisational perspective. The remaining participant had Usability/HCI/IxD/UX as the second field of expertise, with computer security and privacy as main field. The years of experience range from 3 to 28, among 4 female and 3 male. Three of the participants have experience with work in industry as: freelancer consulting on privacy as a competitive advantage, CEO \& head designer for Legal design consultancy, and member of task group of usable security and privacy. All of the participants have academical positions ranging from PhD to Professor. Even though the academic roles are prevalent, we consider these backgrounds to represent well our target group.

The `law group' consists of four people, three having Law/data protection as their main field of expertise. As the second field of expertise one chose again Law/data protection, another chose Certifications/ISO Standards/Regulations, and the other two chose Usability/HCI/IxD/UX. The fourth participant chose Usability/HCI/IxD/UX as primary field of expertise and Law/data protection as secondary expertise. 
The years of experience range from 5 to 14, with 3 female and one male. The balance here is skewed towards academic roles (three out of four) ranging from PhD to Professor, with one participant working for a privacy consultancy firm. For this group it was more difficult to find
people that had also knowledge of usability, besides privacy and data protection.

 \subsection{Data analysis}
 We use thematic analysis (TA) for analysing the data, following \cite{braun2013successful}. We identify the themes in a ``top-down'' fashion, where we use data to explore the concepts of interest (i.e., those mentioned in Section~\ref{subsec_studyQuestions}). Since the analysis is guided by existing theoretical concepts, as well by our standpoints, disciplinary knowledge and epistemology, we adopt a theoretical variant of TA. However, we also employ experential and constructionist variants of TA. For example, a critical and constructionist analysis is used to identify the concepts and ideas that underpin the assumptions and meanings in our data (e.g., we look at how the field of expertise of the participants influences the way they define and understand usability of privacy). We also use TA to develop a detailed descriptive account of usable privacy and related concepts such as processes and criteria for evaluating usable privacy. At the same time, in an experiential TA fashion, we are interested in the participants' standpoints towards evaluating and measuring usability, and how they experience and make sense of privacy/data protection aspects that we define as related to usability.

We adopted a researcher-derived approach while performing our coding. 
When analysing the data, we focused on identifying answers that can be understood as instances that fall (out)inside the concepts that were introduce in \cite{johansen2020making,johansen2020theTR}, i.e.: 
(i) the Usable Privacy Definition, 
(ii) the Usable Privacy Cube model, 
(iii) the Usable Privacy Criteria, and 
(iv) the Usable Privacy Goals. 
Besides these four focus areas, we were also interested in validating our general research question, namely that of adding and integrating usability evaluations into the existing certification schemes. 
The themes have been created based on how meaningful the specific comments of the participants are for the elements we want to validate, how many of the participants have mentioned the specific aspect, but also on how strongly an opinion was articulated and argued for.

\section{Results}\label{sec_results}
The results are presented based on the concepts to be validate from the \cite{johansen2020making,johansen2020theTR}. We start by assessing the need of the certification and standardisation community for evaluating and measuring usability of privacy in Section~\ref{sec_evaluatingandMeasuring}. We continue with validating the UP Definition in Section~\ref{sec_UPDefintion} directly -- by asking the participants in the `usability group' their opinion about the definition we present --, but also indirectly -- by asking all participants, irrespective of group, to present their understanding and experience with usability in privacy / data protection.  In Section~\ref{sec_UPGoals} we validate the list of our UP Goals with the `law group' by asking them to choose the goals that have most relevance for usability, comment on their choices, as well as suggest other items to add to the list. The UP Criteria are validated in Section~\ref{sec_UPCriteria} with the `usability group', where we give additional examples of criteria with their respective sub-criteria, asking the participants to assert whether they represent a good solution for measuring usability of privacy. Finally, as a way of putting all the above concepts together, we present the results from the validation of the UP Cube with the `certification group' in Section~\ref{sec_UPCube}. To conclude, at the end we provide a summary of the overall results and an overview of all themes and their interrelation.

\subsection{The need to evaluate and measure usability of privacy}\label{sec_evaluatingandMeasuring}
\subsubsection{Evaluating usability of privacy}\label{sec_evaluating}
Our general research question -- \emph{``evaluating and measuring on scales the usability of privacy''} -- was formulated in interview questions (Appendix~\ref{appendix_evaluatingMeasuring}) that were addressed specifically to the certifications group, as they are best acquainted with the existing certifications, their needs, and practices. One of the interview questions aimed to elicit whether they find it important to evaluate usability aspects when certifying for compliance with data protection. Their answers are interpreted as \emph{we need evaluations of usability of privacy}. A more detailed look at the answers shows some variations. One of the participants did not use the word ``important'', and sees the evaluation of usability as something that is ``needed'': 
\begin{quote}
``we need evaluations of usability'', ``All the GDPR certification programs or schemas need to also look at usability''  (CertP1).
\end{quote} 
Five of the respondents answered using the word ``important'' with variations such as: 
\begin{quote}``it's very very important\dots'' (CertP6), \end{quote} 
\begin{quote}``indeed is important \dots'' (CertP3), \end{quote} 
\begin{quote}``I do think it's important'' (CertP2), \end{quote} 
while in one of the answers the ``important'' label is not given as a general fact, but as connected to the data subjects' rights: 
\begin{quote}
``it's important also to focus on usability as one aspect related to data subjects' rights'' (CertP5).
\end{quote} 
Moreover, all participants identified several areas where the evaluation of usability is of special importance, or that evaluation should be done ``at least'' in these instances that they exemplified.

One outstanding example (i.e., mentioned by three 
out of four participants that specified cases where usability is important) is that \textit{``usability is important for exercising data subjects' rights''}. Usable transparency and usable intervineability are presented by one of the participants as preconditions for the users to exercise their rights. 
Data Protection Authorities (represented by one of our participants) when evaluating the criteria of certifications schemes for accreditation would 
\begin{quote}
``expect that something also tackling the usability aspects are in place, at least in the case of where transparency is necessary, like how easily understandable is something [pertaining to] Article 13 or 14 [i.e, Information to be provided \dots]. Is it really possible to make a total exercise of your data protection rights, or is it too complicated or not [possible at all]?'' and that ``the end user has to see and understand what is happening'' (CertP1).
\end{quote}
The transparency principle is also mentioned in relation with giving consent: 
\begin{quote}''people need to understand what they're signing or what they are committing to, \dots so that people understand the consent and that consent is not pushed onto them.” (CertP4) 
\end{quote}
The right to be informed is specifically mentioned: 
\begin{quote}``particularly when it comes to the part of the data protection law that talks about information to the user \dots  and actually make them understandable and comprehensive enough for me to make a sort of informed decision” (CertP6). 
\end{quote}
Although not as strongly articulated as the above, data protection by default and by design are also mentioned by 
three respondents 
in relation with evaluating usability in EuroPriSe.  
At this point we can conclude that a sub-theme representative for the `certifications group` is that \begin{quote}
\emph{evaluating usability is important for data subjects to exercise their rights and for data controllers to comply with the transparency principle}.                                                                                                                                                                                                                                            \end{quote} 

\subsubsection{Measuring usability of privacy}

The other side of our general research question -- \emph{``measuring on scales the usability of privacy''} -- is important for making evaluations of usability of privacy more objective and easy to follow by both the companies wanting to be certified and by the certification organizations.
During the interview the respondents were asked whether they see as useful to concretely measure and evaluate how well the usability of privacy is dealt with by companies wishing to be GDPR compliant. We also explained to each participant that by measurements we meant some form of scale or score of the type used to indicate energy consumption for house appliances. The sub-theme that would be representative for the answers at this question is that \emph{Measuring is definitely useful but where do we start?}. That measurements are something desirable was very clearly and strongly stated by all participants. CertP1 even generalized the statement to the whole community: 
\begin{quote}``Yeah, I think measurement is a good thing. It is something everybody or those who are in the community agree on.'' 
\end{quote}
That the community is favorable towards scales based measurements, such as the use of traffic lights, is also exemplified through research work such as \cite{bal2014designing,tesfay2018PrivacyGuide} or by the work done on privacy icons \cite{holtz2011towards,efroni2019privacy}. 

Even though the respondents were in favor of measuring privacy, they all brought up several challenges, and this without us encouraging them to do so.
The concerns being raised are exemplary for indicating were the community is at the moment in terms of measuring (usability of) privacy and what are the challenges, and some possible solutions, that the community sees. Moreover, these answers offer confirmation for the choices taken in \cite{johansen2020making}.

One of the more general challenges that one will be confronted with when measuring usability aspects of privacy comes from the fact that in privacy we do not deal with 
``stabilized knowledge'' 
(according to one of the respondents).
There are examples of actors such as the Stiftung Warentest\footnote{Stiftung Warentest is a German consumer organisation and foundation involved in investigating and comparing goods and services. See \url{https://www.test.de/} .} who compares products/services based on aspects such as usefulness, functionality, or environmental impact, and that are using a scoring system based on percentages (example contributed by one of the respondents). 
However, usability of privacy is not as easy to measure as, for example, the ``consistency for the shampoo'' (CertP1). We have investigated this aspect in \cite{johansen2020theTR}.

One conclusion from several of the participants is that we are still in a rather initial phase regarding measuring usability of privacy, where one still asks basic questions such as: 
\begin{quote}``how do you measure it and what do you measure'' (CertP2)
\end{quote}
or whether
\begin{quote}``is it now a score, is it value, is number \dots or is it only one thing, we need that or not? \dots is there something where we rather need, `This is a no go'?'' (CertP1).
\end{quote}
and we need to 
 \begin{quote}``Come up with a set of objectively measurable aspects of what usability is and what we wish for''. (CertP6)
 \end{quote}

In our work \cite{johansen2020making,johansen2020theTR} we offer a definition of what usable privacy is and we suggest which aspects to evaluate and how to produce measurable results that can be translated later in a number or a score. We do not, however, answer what will be a ``no go'', i.e., what could constitute a minimum level that needs to be reached in order to pass the evaluation. 
Moreover, we propose that usability evaluations should be coming on top of (just) GDPR compliance, with the goal of encouraging companies to improve the usability of privacy of their products (which is being supported also by one of the respondents), 
\begin{quote}``as some type of competition and advantage over others'' (CertP5).
\end{quote}
However, when promoting the `competitive advantage' as a motivation, one has to think of aspects such as 
 \begin{quote}``who's going to pay for the process and it is actually very little happening because of people who can do it, in the sense of the consumer, of the citizen, who usually don't have the resources.'' (CertP4)
 \end{quote}

One continual source of discussion is deciding what can be considered as good enough usability in order to be compliant with data protection regulations, which in turn would define what it means to go beyond, and hence obtain a competitive advantage: 
 \begin{quote}``at some point you would have to discuss if it's still an obligation, a legal obligation, or if maybe some companies go beyond what is required by law'' (CertP5).
 \end{quote} 
In our work \cite{johansen2020theTR} we identify usable privacy goals that appear in the GDPR text, which means that we identify legal requirements for usability of privacy. CertP1 argues that 
 \begin{quote}``as the data protection authority we also have to decide whether it's sufficient, which is proportionate to whether it's OK or not, and sometimes even with the bad usability, it's still okay. It should be better. It could be better, but if we find a system and they only want to know whether they are legally compliant \dots of course changes cost money \dots is it okay or not because we have to decide whether we need to stop data processing because it's unfair, whether we can sanction it or whether something has to be changed. \dots in the end it is about being sufficiently compliant or not? This is where the court has to stop.''
 \end{quote}
 At the same time CertP2 argues that usability
 is 
 \begin{quote}``an element of success, and it's also an element of failure, so it could lead to the failure of compliance. \dots If you make a system that is not usable by the user, then you cannot really have compliance. In the end, it's going to be too difficult, too boring, too bureaucratic, too complicated to do something, and the system will fail.''
 \end{quote}
 The conclusion that we can derive from these discussions is that one has to find \emph{what will constitute enough usability to be deemed compliant and what comes in addition as a competitive advantage.}

Other aspects that were brought up by the participants and that resonate with our work \cite{johansen2020making,johansen2020theTR} are the context of use and the target group:
\begin{quote}``what do you measure in what kind of context and who is the target group''(CertP3);
\end{quote}
\begin{quote} ``Who is the target group of this certification result? Is it the consumer themselves? Or is it maybe an intermediary professional like somebody from a consumer protection organization?''(CertP4);
\end{quote}
 \begin{quote}``we need to know more about the target groups, who is, well, those who have to understand it all” (CertP1).
 \end{quote}
In our Usable Privacy Cube model \cite{johansen2020making} and the Usable Privacy Criteria \cite{johansen2020theTR} we account for the specific context of use, as well as the users with their goals and specific environments. 
The participants also give examples of where the specific context (including the users) is important for determining whether the usability is sufficiently addressed: 
 \begin{quote}``And a very simple example coming to my mind is the number of clicks you need to do something or to get an information\dots because sometimes you can get to control or information in one click, but it doesn't mean that the information is good \dots And sometimes if you have a proper layered approach, maybe you need three clicks, but then the user would be better informed or have better control'' (CertP3)
 \end{quote}
Moreover, an important aspect raised is that different user groups would have different needs regarding the presentation of the evaluation results: 
\begin{quote}``you have certifications or certification results and they are to be looked at primarily by experts or semi-experts like the consultant or the adviser  in a Consumer Protection Organisation. I mean they have some knowledge and they can be able to translate things to their respective target groups.'' (CertP4),
\end{quote}
which is a topic that we cover in depth in the paper \cite{johansen2020privacy}

\subsection{Defining Usable Privacy}\label{sec_UPDefintion}
 In \cite{johansen2020making, johansen2020theTR} we have suggested to adapt the definition of usability from the ISO standard 9241-11:2018 to privacy,
which 
we validate here primarily with experts from HCI/IxD/UX community, as these are supposedly more acquainted with this ISO standard. During the interview we presented the definition and explained how it is relevant for GDPR, after which the respondents from the usability group had to answer whether this definition captures their own (or their community's) current understanding of usable privacy. The multiple-choice answers (i.e., `completely', `partially', `not at all') were followed by an explanation of their choice (Appendix~\ref{appendix_UPDefinition}). 
In addition to asking directly the usability experts to validate our definition, the participants in all three groups have been asked to explain their understanding of usability in the context of data protection, and to also anchor it in the reality of their practice (Appendix~\ref{appendix_perspectiveExperienceOnUsability_q}). 
In order to gather unswayed perspectives, these question were asked in the beginning of the interview, before presenting our definition; in the rest of the paper, we refer to these as `unswayed perspectives on usable privacy'.

From the answers related to our definition of usable privacy we can distill the following theme where all fit:
 \emph{We trust the usability definition from the ISO standard 9241-11:2018}.
The majority of the participants considered the definition as complete, and as an adaptation of the usability definition from the ISO standard, e.g.: 
\begin{quote}
 ``\dots completely because it was almost the same definition of usability [as the ISO standard 9241-11:2018] but trying to have instead of general systems, systems that protect users privacy.'' (UsabilityP1), 
 \end{quote} 
 \begin{quote}
 ``I would say that it is a complete coverage of the different concepts that one could expect within the usable privacy domain because I think indeed there is quite a resemblance to the definition that comes from the ISO standard.'' (UsabilityP2) 
  \end{quote}
 Moreover, besides agreeing with the definition itself, one of the participants also appreciated our exemplification of how the definition applies to GDPR. 
\begin{quote}
``This was a more marvelous thing to see how well you related to the GDPR and to the ISO standard. I think that was wonderful.'' (UsabilityP7)                                                                                                                                                 \end{quote} 
 
We can conclude that all participants agreed that adapting the definition of the ISO 9241-11:2018 to privacy, as it was proposed by us, captures (the choice `completely' being used by the majority, while the remaining chose the alternative `partially') the current understanding of usable privacy in their field (i.e., in usable privacy and security, HCI, UX). 

Further confirmation of this conclusion (i.e., that it is good to base our definition of usable privacy on the ISO standard 9241-11:2018) is provided by two of the answers which are of the type `unswayed perspectives on usable privacy', which mention the definition of usability from the ISO standard when they are asked to present their own understanding of usable privacy: 
\begin{quote}``I can refer to standard definition of usability in this context that they can take control over the data in an efficient way, while they are satisfied and also taking the effectiveness into consideration.'' (UsabilityP1)
\end{quote}
  and 
  \begin{quote}``So I define usability for me and for my students using this ISO definition of usability, where you have effectiveness, efficiency and satisfaction …'' (UsabilityP4)
  \end{quote}
Moreover, these answers provide a good indication that our decision to test the usable privacy definition with the usability group was appropriate, as they are more acquainted with the ISO standard on usability.
 
For the respondents that checked the 'partially` choice, we can group their answers under the theme \emph{Instances of the usable privacy definition}, as these are more specific cases or occurrences of the aspects that are represented at a higher level by the definition. For example, our definition mentions that the goals of the users should be considered, whereas UsabilitP4 says that there should also be a distinction between secondary and primary goals of the users: 
\begin{quote}``Partially because I didn't see this distinguishing between primary and secondary goals. I think it is important. … one could also say that it's some how included, but I think it's better to mention it separately.''
\end{quote}

When validating our definition with the usability experts we have found more such `instances'. One of these is \textit{privacy is a secondary goal to the user}, which can be mapped to ``taking into consideration the particular [goals] of the user'' in our definition. This aspect was mentioned by the UsabilitP4 both when asked to evaluate our definition -- as cited above -- but also in the beginning of the interview, in the `unswayed perspectives on usable privacy': 
\begin{quote}``I consider the three things: effectiveness, efficiency and satisfaction. They should be fulfilled for primary goals and for secondary goals which is in this case data protection. So for example, if some data protection mechanisms interferes with primary goals of the users,\dots too much is of course not good at all, not even if it interferes slightly, or more like middle-ish, I evaluate the things: OK, this can't be usable.''
\end{quote}
Privacy goals as interfering with the primary goals of the users was also mentioned by other participants: 
\begin{quote}``People would be able to do so [understand/read the privacy policies if they are written in a non-legal way], but in practice they don't [read] because it just doesn't work with their lives and it doesn't match the current goal of just signing up for the service and using it.''.
(UsabilityP5)
\end{quote}
\begin{quote}
 ``\dots there are conflicts between giving people enough time to think\dots [to make an informed decision], but they might reasonably have the opinion that they should be able to use something as quickly as possible.''
 (UsabilityP6)
\end{quote}
 \begin{quote}``it's a secondary task, and that means it should be as little, \dots the workload\dots,  should minimise the workload as much as possible.'' 
 (UsabilityP7)
 \end{quote}
Since this topic was brought up by the participants in the usability group only, we mark it as a concern specific to this group. The significance of this topic is also acknowledged by the literature in the field of usable privacy and security, as for example in \cite{whitten1999johnny,acquisti2017nudges}.

In the comment of participant UsabilityP5 above we identify another `instance' of our definition, which we code as \emph{supporting the correct behaviour and characteristics of the users}, which can be mapped to the ``Taking into consideration the particular characteristics of the users \dots'' part of our definition. 
UsabilityP5 argues even more about this point:
\begin{quote}
``I think it should not only be about the ability to access information because you are able to access privacy policies, you're totally able to do that, but nobody does\dots This is like the one of the most traditional things where HCI says yes, okay, formally you provided the information, but practically you didn't reach the goal.''
\end{quote}
Other two participants from the `usability group' argue more lengthy on this topic:
\begin{quote}
 ``you need to understand the behavioural aspects of how people interact with information. Because on one hand, because of information overload, people are [...] using heuristics and proxies like \dots trust, or how famous an organisation is and they just hope that those organisations are doing something legal and reasonable. But people don't really have neither the time nor the willingness to expend effort, many of them, to read what are their rights and what organisations are doing with their data?\\
To actually do disclosures that work for the users, I think that probably usability is also just wider than interacting with the information, because there is still the problem. I believe that even if everyone, every organisation in the world would design these disclosures in a way that it's understandable and user-readable, still, we would need to read and interact with every single one of them, so people would be quite fatigued. \dots So are they standards or [?] so we can guarantee data protection without people having to put any effort?'' (UsabilityP3);
\end{quote}
\begin{quote}
``And people don't necessarily want to understand – some do -- but a lot don't really want to  understand all the details.  They want to know what they have to do and and what they shouldn't do in order to do this correctly.\\
\dots what you see is really just responsibility being pushed down the chain,
and basically exposing people to actually the legal text and various things \dots It's really what I call responsibilization and not really actively enabling and supporting
the correct behaviour.\\
Even if I do not agree, I still have to use it'' (UsabilityP7)
\end{quote}
While two of the `law group' participants mention it as a fact:
\begin{quote}
``Nobody really wants or is able to give an informed consent for every single banner in every single website and application that he or she visits every day.'' (LawP1)
\end{quote}
\begin{quote}
 ``I know that people don't read privacy policies \dots'' (LawP3).
\end{quote}
The literature on usable privacy and security covers well this topic, e.g., speaking of a privacy gap \cite{gerber2018explaining,spiekermann2001privacy} between what the user says that would do when asked or tested in the laboratory and what it actually does when in a concrete, real situation.

Another `instance' that we have identified we call \emph{usability for transparency and data protection rights}. 
This theme includes two elements that appear in the answers of the participant usually as `transparency' and as different aspects of data protection rights such as `having control', `self-determination', `interveneability', `having choices', which are easily correlated.  The certification group, in particular, is preoccupied with both aspects. 
\begin{quote}
 ``The end-user has to see and understand what is happening and we distinguish also between having only the transparency and while having to agree, or also taking steps for your own control; the self-determination, then the possibility to intervene.'' (CertP1)
\end{quote}
\begin{quote}
 ``how easily can they find information on the processing of their data, how easily can they control this and control what kind of data are processed for what kind of purpose, and also how easily they can exercise their rights?'' (CertP2)
\end{quote}
\begin{quote}
 ``I would say that the  primary goal  when it comes to data protection and usability is to create transparency, so that I as a user easily know what  I'm doing and what I'm sending and what these things are used for. So that I actually have knowledge of the data I'm sharing with others and why.'' (CertP6)
\end{quote}
The transparency principle and data subjects' rights, presented intertwined by the `certification group' above, have been identified also in Section~\ref{sec_evaluating} with respect to evaluating usability of privacy.
The `law group' puts more emphasis on the transparency part, in the sense that they bring up challenges, but also solutions, related to how to translate legislative and even technical jargon so that they are more accessible (i.e., as in easy to find, quick to read, and easy to understand) to the regular users.
\begin{quote}
 ``And the difficulty is in really making the translation.'' (LawP2)
\end{quote}
\begin{quote}
 ``\dots almost everything that is in data protection. You know when you try to explain stuff there is a lot of technical jargon,'' (LawP1)
\end{quote}
In addition, the `law group' is committed to suggesting or finding solutions, such as design patterns or privacy enhancing tools, to translate the transparency requirements into usable solutions for the user.
\begin{quote}
 ``Maybe  the way to design concept experiences in which [information] --
  [about] how you protect the data, how you process the data – can actually be designed in a way that makes sense for the user, so that they can make more informed decisions, maybe through, for example design nudges, design friction.'' (LawP1)
\end{quote}
\begin{quote}
 ``I'm discussing ways that we can improve the way that policies are communicated. What I'm trying to do is to implement these policies through the the design or through the interface of the website. So, for example, transparency enhancing mechanisms and or privacy enhancing mechanisms.'' (LawP3)
\end{quote}
Such solutions are being proposed also by participants from the `usability group':
\begin{quote}
 ``we focus on the usable transparency in the consent forms and also the user satisfaction with the affirmative actions and the time that affirmative actions can take from users to handle content forms, and if these affirmative actions, for example can help users to pay more attention, to, for example, make it more accurate for them to remember to do exactly what they gave their consent afterwards.'' (UsabilityP1)
\end{quote}
\begin{quote}
 ``design patterns or some design frictions to help the users to actually get interested in reading'' (UsabilityP3) 
\end{quote}

Another `instance' that was mentioned by many participants, from different groups, is related to `dark patterns' and practices where usability is used against the users. We have coded this instance as \emph{usability turned against users}, which can be mapped to our whole definition, albeit negated, as follows: ``Dark usable privacy refers to the extent to which a product or a service deceives the users to give up their privacy in an efficient, effective, and illusory/seemingly satisfactory way, by taking into consideration \dots''. 
\begin{quote}
 ``There's actually the opposite of usability or anyway usability being used for dark patterns. So design expertise being willfully used by big ones. So organisation that definitely know what good UX is like Facebook or Google that are thwarting the choices of users making it more difficult to change something, or to say no, you cannot track me and so on. So actually there is quite sophisticated work, but done against consumers.'' (UsabilityP3)
\end{quote}

The above `instances' follow a pattern that we already have used when doing the work presented in \cite{johansen2020making,johansen2020theTR}, i.e., that of finding out what are the usability aspects that need to be addressed and then find solutions for dealing with them. 
Our approach to identifying `instances' of usable privacy was to go through the GDPR text and mark all the  cases that could be considered vague and interpretable, and if evaluated in usability tests could result in different levels of achievement of the usability goals: effectiveness, efficiency, and satisfaction. In the paper \cite{johansen2020making,johansen2020theTR} we identified most occurrences in the provisions concerning consent, then on the second and third place, provisions regarding the transparency principle, and respectively the rights of the data subjects. We can see here an overlap between the `instances' we ranged on the second and third place in \cite{johansen2020making,johansen2020theTR} and the ones that we surfaced in the analysis done in this section, which we called  \emph{usability for transparency and data protection rights}. 

\subsection{A comprehensive list of Usable Privacy Goals}\label{sec_UPGoals}

In the previous section we identified aspects of privacy and data protection for which usability is considered relevant by our participants, with differences between the three groups becoming apparent. 
What we called Usable Privacy (UP) Goals in \cite{johansen2020making,johansen2020theTR} can also be considered `instances' of the usable privacy definition, in the same sense as above. 
Here we validate our UP goals with the `law group', since this is well acquainted with the GDPR text, from which we have extracted our UP Goals. 

The participants in the `law group' were given a list with all our UP Goals (see Appendix~\ref{appendix_UPGoals}) and were asked to choose the ones that they thought relate to usability. We then discussed their choices and opinion about this list, whether they thought it was exhaustive, and whether they could provide additional goals. 

Counting the numbers of goals from the list checked by the participants, the mean is 21,75 choices out of 28, i.e., a 77,67\% coverage.  
Thus, the participants generally agree with our UP Goals, where particularly LawP1 checked all the goals, whereas LawP3 and LawP4 expressed directly their satisfaction with how well the list covers usability aspects:
\begin{quote}
``\dots your list was very complete. I cannot think of something that is not on this list. \dots I think this list here is very broad and very comprehensive regarding usability. I cannot think of anything else.'' (LawP3)
\end{quote}

\begin{quote}
 ``I am happy with this list. I'm extremely happy with this list.'' (LawP4)
\end{quote}

However, the same two participants also said that if they would go thorough the GDPR text they will probably find more such goals, but they could not name any such goals in the interview.

\begin{quote}
 ``I think if I had more time to think and analyse and probably go back to GDPR I would find other usability goals.'' (LawP3)
\end{quote}

\begin{quote}
 ``I don't think it's exhaustive, but I would need to scroll all the GDPR.'' (LawP4)
\end{quote}

Therefore, from these comments and the large number of UP Goals selected (i.e., confirmed) we can derive the following theme: \emph{I am happy with the list of Usable Privacy Goals}.  

Analysing closer the not-checked UP Goals reveals uncertain UP goals since a direct relation to usability is not obvious.
The UP Goal ``Make the natural persons aware of risks, rules, safeguards and rights in relation to the processing of personal data. [Recital (39) of GDPR]'' was not checked by LawP2 and LawP3. While having no explanation from LawP2, we can see that LawP3 associates this with other aspects than usability:
\begin{quote}
 ``it's more related to language, and I tend to not associate so much the language with usability.''
\end{quote}
Even LawP1 shows initial hesitation towards this goal, but agrees with it in the end:
\begin{quote}
``I'm not sure about how awareness goes hand in hand with usability. It's an important ingredient, is an important aspect but I'm not sure whether one determines the other. \dots But awareness of risks it's different. I think it's important. Awareness of the extent to which consent is given; I know it's a usability issue.''
\end{quote}
However, opposed to these three participants, LawP4 is strong on including this goal, together with ``neutrality of privacy choices'' and even arguments with literature references \cite{feng2021aDesignSpace}.
\begin{quote}
 ``I mentioned in the beginning like the user awareness of the risks, of the of the consequences, but I see here Recital (39) being quote, being here. So I'm very happy with this. \dots 
The neutrality aspect it's not here. I think neutrality is very important. It's also I think very related to usability.  Actually, it's one of the elements of the design of privacy choices, of the method that I mentioned from [Feng and Yao].''
\end{quote}

Other goals that need further scrutiny, being left out by three of the participants are: 
\begin{description}
\item[(UPG.11)] ``Personal data should be processed only if the purpose of the processing could not reasonably be fulfilled by other means. [Recital (39) of GDPR]'' and
\item[(UPG.30)] ``The data subject should have the right not to be subject to a decision based solely on automated processing, including profiling, which produces legal effects concerning him or her or similarly significantly affects him or her. [Article 22 (1) of GDPR]''
\end{description}
Despite not being given reasons for why the three participants did not choose the UPG.11, the UPG.30 is checked and argued for by the LawP1 with the following comment:
\begin{quote}
 ``And also the last one [number 28 in the list], I think there should be a very easy actionable choice to not be subject to decisions based on automated processing.''
\end{quote}

LawP2 gives a more detailed account of the choices made in the list, where particularly interesting is 
mentioning
`control' as the more general area, while goals related to consent and data subject rights are seen as more specific. Since in \cite{johansen2020making,johansen2020theTR} we decide the level of achievement for a ``generic'' goals based on measurements coming from more specific criteria, it is also interesting to observe how the participant correlates consent and data subject rights to control:
\begin{quote}
 ``\dots more in general I have selected a number of sentences which relate to the control aspects \dots, but also the specific informed indication of your agreement. So about consent.''
\end{quote}
and about how consent relates to data subject rights:
\begin{quote}
 ``having control of your personal data, \dots that's really about being able to decide for yourself which data you're sharing, having insights in these data. This also relates to the exercise of data subject rights,  having access to your data -- obtaining this access -- in an easy way,  being able to check whether things are correct indeed, and also having the option to object.''
\end{quote}

Indicating such general usability related areas matches the approach we adopted in \cite{johansen2020making,johansen2020theTR} where the UP Goals and UP Criteria (see also Section~\ref{sec_UPCriteria}) are categorized based on their area of application from the GDPR text. The areas listed below are ranged based on the number of goals found in each category, the first item having the most occurrences. 
\begin{enumerate}
\item Consent (lawful grounds for processing data principle),
\item Information and communication addressed to the public or to the data subject (transparency principle),
\item Rights of the data subjects (rights in general),
\item Purpose of processing, and
\item Legitimate interest of either the processor or the data subject (lawful grounds for processing data principle).
\end{enumerate}
In addition to areas related to preexisting GDPR chapters, we also have a ``generic'' or higher level category, where we have placed UPG.2: ``Natural persons should have control of their own personal data. [Recital (7) of GDPR]''.

``Having control'' is the aspect that is valued highest by LawP2 in respect to usability. Although control is only mentioned in Recital (7) of GDPR, and mostly relates to consent
\begin{quote}
 ``It's difficult to get it from the GDPR, except for the Recital (7) I was mentioning. \dots I think most control and most usability is related to situations where you are indeed basing the processing activities on informed consent. But then it should not only be about being informed, having a proper choice and really knowing what you're consenting to.'' (LawP2)
\end{quote}
 LawP2 sees having control on which data one is sharing as the next level in usability of privacy:
 \begin{quote}
``But the next step for me would really be to have more control and more fine grained control in which data then you are sharing,
for which purposes and making more fine grained selections of how the data are used by a data controller. At least it would really be the next level on usability of privacy.''
 \end{quote}

\subsection{Ways to meet the Usable Privacy Criteria}\label{sec_UPCriteria}

Having established the list of usability goals 
that GDPR stipulates,
the practice in the Interaction Design field is to operationalize these by turning them into usability criteria formulated as questions  \cite{preece2015interaction}. Criteria can be seen as specific objectives to be reached by those that aim to reach the set of goals that the criteria relate to. In our case, the Usable Privacy Criteria enable one to asses the privacy related measures that a product or system provides in terms of how much these improve the control that the data subjects have over their data. Examples of commonly used usability criteria (i.e., not specific to only privacy) are:
\begin{enumerate}[(i)]
 \item time used to complete a task (efficiency), such as reading a privacy statement, or 
 \item the number of errors made when carrying out a given task (effectiveness), such as choosing the desired privacy settings.
\end{enumerate}
Usability criteria can provide quantitative indicators of the extent to which, for example, the data subjects understand the implications for her/his privacy from using a certain technology. 
In addition, given that certification bodies already use various forms of criteria in their evaluations, we too want to provide the Usable Privacy Criteria as a set of rules for assessing also usability, so to enable the certification bodies to integrate evaluations of usability within their existing certification schemes.

The UP Criteria are validated in this study with the `usability group', as they are most acquainted with the process of formulating criteria to meet goals such as efficiency, effectiveness, and satisfaction. 

The participants were given examples of the UP Criteria and were asked to comment on them (see Annex~\ref{appendix_UPCriteria}). The UP Criteria have been assessed as good by most participants, using quick and simple statements, such as:
\begin{quote}
``I think they are a good solution for evaluating usability.'' (UsabilityP1)\\
``I definitely see the reasoning behind it and it makes sense for me.'' (UsabilityP2)\\
``The questions are good to me.'' (UsabilityP3)\\
``I think it's good that you  have these measurement tools or dimensions so that you can start to negotiate how you should measure all these individual questions.''  (UsabilityP5)\\
 ``Yeah, I think the criteria are good.'' (UsabilityP6)
\end{quote}

However, the participants were keen on the discussion to quickly turn towards another related topic that seams to be preoccupying the community at the moment, that of establishing standards, recommendations, and creating guidelines or design patterns, to help with meeting such criteria, e.g.:
\begin{quote}
``probably giving some sort of rubrics or somehow like figuring out what are the  things that the organisation can possibly do, for example to give control to people of their personal data. You design a rubric with scenarios like is there a dashboard? Yes. Is the dashboard clear? Yes. Are all the buttons equal and there's no dark patterns? And then you start scoring on a more grained way like that. Perhaps you can help also non-designers to evaluate these things or you can help the actors that would like to do it, but they don't have a clue how to do it.'' (UsabilityP3)
\end{quote}
\begin{quote}
 ``To measure understanding of the consent text, you should do this. Whatever it is. For example, you should have to understand the questions about this, or you should have to make user interviews and see what users think when they read this text. Yeah, so something like concrete. What concretely to do in order to understand if the user understands?'' (UsabilityP4)
\end{quote}
\begin{quote}
``there is a  Article 29 Working Party opinion on purpose limitation and implementation of the principle of purpose limitation and there they say a little bit more, or at least this could be a guideline for implementing information and unambiguous consent.'' (UsabilityP5)
\end{quote}
\begin{quote}
`` your questions will really be questions to start with: How long, how much, et cetera? And then you start to see the patterns that are leading to good informed consent and then you say that, well, we take these as our standards now. \dots I would say that within a branch of industry or branch of services, you could say that after a while you will get established. You know this is sufficient for this kind of service. This kind of data request has become the standard.
I think that the Data Protection Board or Consumer Agency are doing such evaluations.'' (UsabilityP6)
\end{quote}

The general tendency to transform the usability aspects into very concrete privacy practice is reflected also in the law domain, as directly expressed by one of our participants from the `law group'.
\begin{quote}
``I see a lot of development of usability aspects / usability concepts, how some principles of usability, if I can say so, are decomposing to very concrete 
privacy practises for data controllers and for users, so indeed there is a prolongation, there is some decomposition, some instantiation of it and not only at the hard law, but also on soft low level.  \dots encompassing usability aspects mostly at the guideline level. So you will not see examples of icons in the GDPR, but you will see a lot in DPA guidelines. References to dashboards, or what does it mean intelligibility or clear and plain language you will see it in the guidelines, you will see it decomposed in case law. \dots there is some very important decision from that CNIL versus Google in 2019 that talks about ergonomics of information not only informed consent or informed choices, but how the information should be delivered.'' (LawP4)
\end{quote}

That such more concrete guidance is needed is confirmed by the participants in the `certifications group' as well. Their assent is especially valuable as they are the ones that are actually performing the evaluation in practice. 
\begin{quote}
``but it would need, for me, a very specific grading system that an auditor could check the effectiveness and the efficiency in the use of usability [against] because these terms are very experienced based, they're very subjective, right? It's not like if you're asking them: Is the colour blue? You're asking if it's usable and they have to compare it to something and say it is usable. So unless you provide a very concrete scheme behind this, I don't know how if it's going to work.'' (CertP2) 
\end{quote}

The focus of the participants was on the particularities of the evaluation, addressing questions such as who would perform the evaluation, what kind of expertise the evaluators would need to have, or which specific HCI methods should they use. 

\begin{quote}
``What type of the methods should I use to evaluate this and then what type of users, the number of the users \dots'' (UsabilityP1)
\end{quote}

\begin{quote}
``So this kind of guidance that you have here would be very useful for somebody who is professional in both usability and privacy \dots I could imagine that people who are experts in only one, or people who are experts in law  would need a lot of guidance \dots'' (UsabilityP4)
\end{quote}

Since the UP Criteria functioned more as a trigger for discussing other more particular aspects of the evaluation process, a theme that would characterize best the type of feedback we received from the participants is \emph{Ways to meet the Usable Privacy Criteria}.

One of the participants from the `certifications group' indicates nicely where we are at the moment with this endeavor.
\begin{quote}
``There's some guidance right now, [and] I think it will be refined more and more. Still it won't be on the same level as: Your sentences can only be half this length, or something which can also be automated. On the other hand, I think we may have design patterns which are okay, [as for example] some grouping like you use for content management, and this [will be] approved by data protection authorities or by court, and then perhaps 20 years later people think: Let's rephrase it a bit. Now it's even better. \dots On the other hand, we are not there yet. If you see the cookie banners there's so many different ways to give consent or [to present the information for the data subjects] to understand what is essential.'' (UsabilityP1)
\end{quote}

\subsection{Usable Privacy Model -- an abstract representation of known and implied principles of privacy / data protection evaluations}\label{sec_UPCube}

Once having proposed a set of criteria, 
we are further interested in 
how these would be integrated in the processes and with the criteria of evaluation of the existing certification bodies. 
Particularly, \cite{johansen2020making} studies the certification scheme of EuroPriSe, and devises a higher level model, called the Usable Privacy Cube (or UP Cube), that combines the evaluation criteria of EuroPriSe with the Usable Privacy Criteria. 
Building the Usable Privacy Cube model on top of the EuroPriSe criteria gives a good anchoring in this well established privacy certification scheme, but even more, it provides a general reorganization of the EuroPriSe criteria into rights and principles as found in GDPR, which is meant to work as a guide for adding usability evaluations to other data protection certifications that target GDPR. 

In this study we validate whether the UP Cube model reflects the existing privacy and data protection evaluation processes, and to what extend (i.e., totally or partially). Specifically, we discuss with the participants the following features of the UP Cube model:  
\begin{enumerate}[(i)]
\item representing the perspectives of both data subjects and controllers/processors; 
\item grouping, prioritization, and organization of the criteria; 
\item interactions between the different criteria; and 
\item context of use (or context of processing, as a term often used in GDPR).                                                                                                                                                                                                               \end{enumerate}

To our question ``Does the UP Cube model represent, at a high level, the existing data protection and privacy evaluation processes?'', two out of the five participant chose ``Completely'', while three chose ``Partially''. 

The answer of the CertP1 is exemplary: 
\begin{quote}
 ``What I know best is EuroPriSe and the previous data protection seals from ULD [Landeszentrum für Datenschutz Schleswig-Holstein\footnote{\url{https://www.datenschutzzentrum.de/guetesiegel/}}], so it's quite very much related, but I think not completely. So I would say partially, although on the abstract level will be the same as the Standard Data Protection model that also uses the different axes for something like that. So the general principle I think is quite well known \dots''
\end{quote}
The Standard Data Protection model \footnote{\url{https://www.datenschutzzentrum.de/uploads/sdm/SDM-Methodology_V1.0.pdf}} has the notion of allocating the legal requirements of the German Federal Data Protection Act -- BDGS (Data minimisation, Availability, Integrity, Confidentiality, Unlinkability, Transparency, Intervenability) to the protection goals, in a tabular manner. 
A cube (like our model) can be understood as a three-dimensional tabulation mechanism, i.e., represents three tables, each based on the combination of two of the axes. Therefore, the model mentioned by the respondent can replace the EuroPriSe in the base square of our UP Cube, and it is already fitting, to some extent, withing our two axes of organization in rights and principles. This example also confirms the versatility of the UP Cube model.

Moreover, CertP2 confirms the novelty and usefulness of the UP Cube model as it makes clear certification aspects that are not explicitly expressed anywhere, but more implied:
\begin{quote}
``I would say that it's something that's not actually ever written down, that's more like implied. The auditors correlate these things when they're doing the audit, but there's nothing written down that says that they should do it like this. Yeah, so I would say partially, very abstract, actually.''
\end{quote}

What is more interesting is that, even though CertP5 and CertP6 chose ``Completely'', their answers are on the same line with the first two participants, in the sense that the ideas behind the UP Cube model are known, 
and that it represents these at an abstract level (as a model is meant to do). 
Therefore, the theme that we extract from these answers is that 
\emph{``UP Cube is an abstract representation of known, but implied or covert practices''}.

\begin{quote}
``Yes, probably your model is representing everything at a more abstract level because it's probably covering all four aspects. So I could say completely, but sometimes not very specific.'' (CertP4)
\end{quote}

As in the case of the Usable Privacy Definition presented in Section~\ref{sec_UPDefintion}, we had questions preceding the presentation of our model (these are called the ``unswayed perspectives''), asking if the participants know whether the certifications or standards that they are acquainted with have a high-level model to guide the process of evaluation. The conclusion from these answers is that it does not exists a published or well established model to guide the process of the evaluation, but there are some main guiding pillars. These are following the GDPR text, or in the case of the standards for evaluating management systems, the risk management or the Privacy Impact Analysis (PIA) is the focal point.
\begin{quote}
``For the time being the guidance for certification against 27701\footnote{ISO/IEC 27701:2019 Security techniques -- Extension to ISO/IEC 27001 and ISO/IEC 27002 for privacy information management -- Requirements and guidelines, \url{https://www.iso.org/standard/71670.html}} is very generic.
So you're not going to find something that would be concrete for the evaluation process, so it's just the they give you (?) with the requirements. They tell you that the organization should do these things and then it is asked to evaluate based on understanding and experience whether they achieve the goals of its control.
They do not provide any more guidance than that. So they tell you, they [the organizations] should have a record of processing facilities, for example, this record could be in Excel, it could be a word file, it could be some paper that they write down. It could be anything as long as they have a record of processing facilities; from an audit point of view this is okay?'' (CertP2)
\end{quote}

\begin{quote}
``So what I know about the former and also the present certification schemes is that they are driven by the legal text and driven by additional criteria, perhaps from the Article 29 Working Group \dots'' (CertP1)
\end{quote}

\begin{quote}
``another tool that isn't GDPR is the privacy impact assessment.
This is a way to evaluate processing in regard of the risks that can happen or that are linked to it.'' (CertP3)
\end{quote}

\begin{quote}
``The ones I primarily know of are management system standards and \dots they are also risk based, which means that you choose measures based on your perceived risk.'' (CertP6)
\end{quote}

When discussing with the participants the individual components of our model, 
we obtain further confirmation of our previous
observation that the elements of our model are present, to a certain degree, in the existing processes.
Participants confirm that even elements that we deem indispensable for a usability evaluation, such as looking at the context of use or the different stakeholders, are also being considered to a certain extent in some of the certification schemes. 

We next give an overview of the level of the consideration given at the moment to aspects related to usability evaluation in privacy certification. This overview can indicate the premises one would have to start from when 
considering evaluating to what extent products or services meet the usability goals of GDPR, identified by us in \cite{johansen2020theTR}.

The idea that the UP Cube model should capture both the perspective of the data subjects and of the controllers/processor, resonates well with our participants.
\begin{quote}
 ``It is smart to have these different axes so that you can see at one glance
how things are related to each other. We have that in our criteria catalogue. For example, we would have the legal obligations of, let's say the controller regarding information duties and on the other hand we would also go to the other perspective of the data subject and we would have another requirement [related to the] right to be informed. So we have these connections also. But they are not that visible of course as in the case of the cube. So I think that this is one of the benefits that you would really present immediately to the interested parties, and I think the cube would work for both. That's your intention. It would bring benefit to the data subjects and also to the controllers \dots
I think that's a good approach to do it that way.'' (UsabilityP5)
\end{quote}

We generally see in the discourse of our participants, irrespective of group -- when discussing about usability in general or other topics not related to the UP Cube model -- that they are preoccupied with both perspectives. In the case of the controllers, usability is considered as the ease with which they can implement data protection related requirements and become compliant.

\begin{quote}
So usability is always a battle in IT, we have to make sure, especially in IT security and of course in privacy, that whatever we create can be used by the people that are supposed to implement the specific activities without creating too much of a burden. Especially the early systems that we saw in compliance to either information security or data protection they used to be very paper based and they were very bureaucratic and they created a burden to anybody that had to perform these actions and especially the parts where they had to ask for specific permissions or consent, or they had to monitor the information that they were using through the record of processing activities. So for me, usability is to achieve your goals regarding information security and privacy, data privacy, while also allowing another person to perform their tasks seemingly unaffected by the fact that they are being compliant.'' (UsabilitP2)
\end{quote}

\begin{quote}
``I'm thinking about end users and how data privacy might be made more usable and a better user experience for them. But I'm also thinking about what practitioners and researchers have to do to deal with data protection issues, myself included, \dots we have been running a lot of projects internally in which we gather personal data, even sensitive data. So we have to navigate the complexities of data protection compliance for example, which is really unusable. It's not a very good user experience for anybody as researcher, especially because you have a lot of obligations upon you, not only in better data protection but also in terms of ethical issues, in terms of how you manage data in general, not only personal data, but how do you protect and manage the data that you produce through the research. Of course is also a little bit broader, but then a core part of it is the personal data that you manage an and it's very very very complex, \dots for me it's also research question: How can we make data protection compliance usable?'' (LawP1)
\end{quote}

In practice, the rights and obligations of the parts involved need to be balanced, and the balance is achieved through a process of negotiation, where considering the specific context of data processing is seen as indispensable.  

\begin{quote}
``what we also see is that very often there are some rights of the different parties, and \dots one is always already in a better power position. So usually the data controller.
We would assume they are responsible for the personal data, they are the power and they have to protect the personal data of the data subjects \dots then the question is what are then the rights for each of the of the [parts involved]. Sometimes people who haven't paid their their bills claiming that the data has to be deleted because they are not customer anymore. They haven't paid, but they are not customers anymore. Now everything because of data protection laws has to be deleted. And obviously there is the right of the company to have paid invoice \dots We also see similar things with the employer employee relation, where the employer should not be surveyed 24/7 or should have some personal freedoms. But of course must not misuse the resources of the employer of the organisation of the company. So these rights always have to be discussed an also weighed so that proportionate solution are found \dots  the situation for an online shop is usually different from the employee employer situation or from a police control situation and therefore we very often have to keep in mind what is the problem behind [these different situations] and  how to leverage the different positions.'' (CertP1)
\end{quote}

From the examples above we see that it is necessary to consider the specificities of different ``situations'' being evaluated, in order to achieve the optimal power balance and protect the rights and interests of all parties involved. Similarly, we see that in the case of all existing evaluations the context of use/processing is considered in relation to, for example, the risk analysis or the type of data being processed, but this is not done in an explicit way as a usability evaluation would require.

\begin{quote}
``Not as such. I would say that's usually implicitly added into the risk management model,  as GDPR as well \dots'' (CertP6)                                                                                                                            \end{quote} 

\begin{quote}
``So we definitely considered it from the point of view of the data and the criticality of the data, but we go with the categories of data, we don't go the with the the context.
You look at what type of information is being processed and then you are judging whether it's suitable for the use, whether the principle of data minimization for example and all the other principles are being imposed for that type of information.'' (CertP2)                                                                                                                                                                                                                                                                  \end{quote} 

\begin{quote}
``In terms of talking about let's say a product or also a processor service, as we call it, then it's of course important who is addressed by that service. Is it, for example, exclusively offered to, let's say lawyers or doctors, than you would of course have to consider: Since it's a lawyer, I would mainly first of all consider legal specifics in a certain setting. But you could also of course go beyond that, you have to understand how the specific sector in which something is used, how this context works, because otherwise you you cannot really go for certification there. So you need at least a basic understanding of that. So these are definitely aspects that would matter here and in terms of the context of processing also and then it's getting related somehow to the risk that is associated to the processing. This of course also important in terms of applying our criteria and principles an it's somehow similar probably, but also different if the data subjects would be the users;  then of course it's once again important to know, are they patients who want to make use of some, let's say health app or some internet portal.'' (CertP5).                                                                                                                                                                                                                                                                                                                                                                                                                                                                                                                                                                                                                                                                                                                                                                                                                                                                                                                                                                                                                                                                                                                                                                                                \end{quote}

\subsection{Discussion}

\begin{figure*}[t]
\centering
\includegraphics[width=\textwidth]{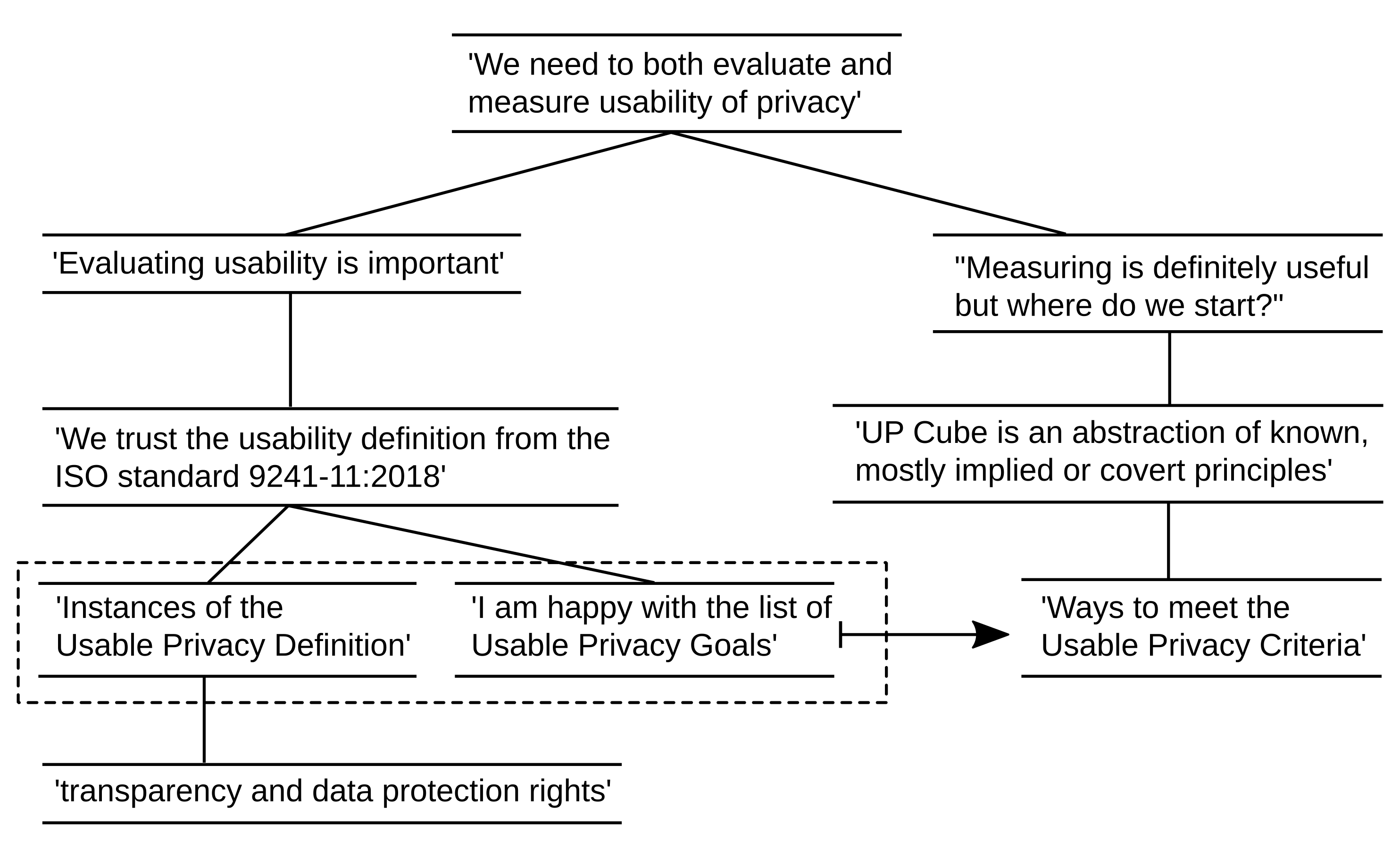}
\caption{Overview of hierarchical and lateral themes and how they relate to our research concepts: Usable Privacy Cube, Definition, Goals, and Criteria.} 
\label{fig_themesMap}
\end{figure*}

We present in Figure~\ref{fig_themesMap} an overview of the themes and their sub-themes that were identified in the analysis above. The themes are presented hierarchically and in relation with the concepts from \cite{johansen2020making} that are being evaluated. At the top of the hierarchy we have a theme that is closest related to our main research question. This is a confirmation of the importance of the topic we approach in our research, namely \emph{``We need to both evaluate and measure usability of privacy''}. 

Looking closer at both the ``evaluation'' and ``measuring'' parts we see that the participants are preoccupied with finding out how to go about measuring -- \emph{``Measuring is definitely useful but where do we start?''} -- and also what are the privacy aspects that are relevant to usability and usability evaluations. \emph{``Transparency and data protection rights''} seems to be the motif that appears mentioned in several context, for example as something important to measure, or as an \emph{``Instance of the usable privacy definition''}. The themes related to \emph{``Transparency and data protection rights''} overlap with the categories that we form for our Usable Privacy Goals and Usable Privacy Criteria, both being evaluated favorably by our participants. We conclude that this is an important area to consider and prioritize when approaching these research question. 

We see one other overlapping area between our research in \cite{johansen2020making} and the answers of the participants, which is the tendency to identify `instances' of usability relevant for privacy. The Usable Privacy Goals represent instances of usability aspects that we identified in the GDPR text, while the participants have been pointed to such instances both when asked to validate our Usable Privacy Definition, as well as when they presented their understanding of usable privacy/data protection. 

The theme that we present at the top of the hierarchy \emph{``Measuring is definitely useful but where do we start?''} resurfaces in relation to our UP Criteria, as the participants are preoccupied with finding and defining concrete ways of meeting the Usable Privacy Criteria, such as which exact HCI methods to use in the evaluation.  

The Usable Privacy Cube is a model that brings together all our other concepts and shows how one could integrate evaluations of privacy in the existing certification schemes. The Cube is also our solution to how to approach the evaluation of privacy. Hence the grey box that encloses the UP Definition, Goals, and Criteria, and the arrow pointing from the \emph{``Measuring is definitely useful but where do we start''} to the UP Cube.  We have validated our assumption that indeed the concepts in the model represent at a more abstract level elements that are already considered by the auditors in their evaluations. However, these are mostly ``implied and covert principles'', and especially those that are important for usability have often little influence on the final appraisal.

\section{Conclusion}\label{sec_conclusion}

In the present study we have validated concepts that were introduced in our paper \cite{johansen2020making}, namely: 
(i) a Definition of Usable Privacy, 
(ii) a set of 30 Usable Privacy Goals 
(which can be seen as instances of the Usable Privacy definition), 
(iii) Usable Privacy Criteria, and 
(iv) the Usable Privacy Cube model. 
These concepts have been introduced as the first steps towards creating a methodology for evaluating, and measuring on scales, usability related aspects of the data protection law.  
We interviewed experts from three relevant fields of practice and research: data protection law, privacy/data protection certifications and standardization, and usability (spanning fields such as Human-Computer Interaction, Usable Privacy and Security, or User Experience). The experts were asked to express their knowledge, understanding, and/or opinions on the above four concepts/topics. 

The study plan used one group of experts to address one specific topic; the expertise of the group was thus thought to match the topic. However, we did not held interviews with focus groups, but with the individual participants. Therefore, the analysis of each of the four topics is done within the frame of one group. Nevertheless, the topic of usability in privacy being more general was addressed by all participants and was therefore analysed across the groups. Moreover, we sometimes found answers in one group to be relevant for another topic than the one in focus. We thus often use such additional opinions to strengthen the findings within a group. 

A second design aspect of the study was to have two main parts: 
one where the interviewees present their opinions without being influenced by our research, and 
another where we present to them our research after which we ask them to directly comment on what was presented. The answers from the first part were used often to corroborate the responses after the presentation, and we found that the participants were consequent with their opinions, the change being only in adapting their answers to what was relevant for our research topics.

The \emph{limitations} of our study come from the fact that though we initially intended to have an equal number of participants for each group, it was a challenge to find people with a background in law that would also have a decent level of understanding of usability matters. One solution that we can see for future work is to organize workshops where we recruit participants with only knowledge of digital privacy and data protection for IT systems and use more time to introduce them to usability related theory. In addition, as we use an interpretative framework based on pragmatism, we wished for a better balance than we achieved between the industry and academic backgrounds in the case of the `law' and `usability' groups. 

The opinions expressed in the interviews are substantially endorsing our research approach, both directly and indirectly. We give an example of an indirect confirmation: 
when the `usability group` was asked to validate our definition of usable privacy, after agreeing in majority with our approach, they suggest more specific aspects to be taken in consideration. Looking for which are the specific areas of data protection that usability is relevant for, is what we do when compiling our list with Usable Privacy Goals. The analysis shows that there are overlaps between the UP goals and their groupings and the `instances' proposed by the participants. One such prominent example being \emph{transparency and data protection rights}. 

A characteristic of all the participants was to look for more concrete, particular, and practical aspects to address in the future and to suggest possible solutions. For example, in conjunction with validating the UP Criteria, they were pointing out what is yet to be done to meet these criteria, and even proposing possible solutions, as for example: ``\dots some sort of notes from reviewers or tips. Perhaps even -- let's say that this would be a formal evaluation -- saying in advance what is it that you're looking for, so people can start thinking: Okay, if these are the things that get scored, what are the practises that are considered good so that we can start doing them on our own?'' (UsabilityP3). Some of these overlap with what we propose as further work in our paper \cite{johansen2020making}, as giving examples of which HCI methods to use in the evaluation. Other important aspects mentioned as necessary in order to help the organizations meet these criteria was to give them some design patterns to follow, or example of best practices. These answers constitute a good list of open problems that the community can address.

%

\begin{thebibliography}{19}
\expandafter\ifx\csname natexlab\endcsname\relax\def\natexlab#1{#1}\fi
\providecommand{\url}[1]{\texttt{#1}}
\providecommand{\href}[2]{#2}
\providecommand{\path}[1]{#1}
\providecommand{\DOIprefix}{doi:}
\providecommand{\ArXivprefix}{arXiv:}
\providecommand{\URLprefix}{URL: }
\providecommand{\Pubmedprefix}{pmid:}
\providecommand{\doi}[1]{\href{http://dx.doi.org/#1}{\path{#1}}}
\providecommand{\Pubmed}[1]{\href{pmid:#1}{\path{#1}}}
\providecommand{\bibinfo}[2]{#2}
\ifx\xfnm\relax \def\xfnm[#1]{\unskip,\space#1}\fi
\bibitem[{Acquisti et~al.(2017)Acquisti, Adjerid, Balebako, Brandimarte,
  Cranor, Komanduri, Leon, Sadeh, Schaub, Sleeper, Wang and
  Wilson}]{acquisti2017nudges}
\bibinfo{author}{Acquisti, A.}, \bibinfo{author}{Adjerid, I.},
  \bibinfo{author}{Balebako, R.}, \bibinfo{author}{Brandimarte, L.},
  \bibinfo{author}{Cranor, L.F.}, \bibinfo{author}{Komanduri, S.},
  \bibinfo{author}{Leon, P.G.}, \bibinfo{author}{Sadeh, N.},
  \bibinfo{author}{Schaub, F.}, \bibinfo{author}{Sleeper, M.},
  \bibinfo{author}{Wang, Y.}, \bibinfo{author}{Wilson, S.},
  \bibinfo{year}{2017}.
\newblock \bibinfo{title}{{Nudges for Privacy and Security: Understanding and
  Assisting Users’ Choices Online}}.
\newblock \bibinfo{journal}{ACM Computing Surveys (CSUR)} \bibinfo{volume}{50},
  \bibinfo{pages}{1--41}.
\bibitem[{Angen(2000)}]{angen2000evaluating}
\bibinfo{author}{Angen, M.J.}, \bibinfo{year}{2000}.
\newblock \bibinfo{title}{{Evaluating Interpretive Inquiry: Reviewing the
  Validity Debate and Opening the Dialogue}}.
\newblock \bibinfo{journal}{Qualitative health research} \bibinfo{volume}{10},
  \bibinfo{pages}{378--395}.
\bibitem[{Bal(2014)}]{bal2014designing}
\bibinfo{author}{Bal, G.}, \bibinfo{year}{2014}.
\newblock \bibinfo{title}{{Designing Privacy Indicators for Smartphone App
  Markets: A New Perspective on the Nature of Privacy Risks of Apps}}, in:
  \bibinfo{booktitle}{{Americas Conference on Information Systems (AMCIS)}}.
\bibitem[{Braun and Clarke(2013)}]{braun2013successful}
\bibinfo{author}{Braun, V.}, \bibinfo{author}{Clarke, V.},
  \bibinfo{year}{2013}.
\newblock \bibinfo{title}{Successful qualitative research: A practical guide
  for beginners}.
\newblock \bibinfo{publisher}{Sage}.
\bibitem[{Efroni et~al.(2019)Efroni, Metzger, Mischau and
  Schirmbeck}]{efroni2019privacy}
\bibinfo{author}{Efroni, Z.}, \bibinfo{author}{Metzger, J.},
  \bibinfo{author}{Mischau, L.}, \bibinfo{author}{Schirmbeck, M.},
  \bibinfo{year}{2019}.
\newblock \bibinfo{title}{{Privacy Icons: A Risk-Based Approach to
  Visualisation of Data Processing}}.
\newblock \bibinfo{journal}{European Data Protection Law Review}
  \bibinfo{volume}{5}, \bibinfo{pages}{352--366}.
\newblock \DOIprefix\doi{10.21552/edpl/2019/3/9}.
\bibitem[{Feng et~al.(2021)Feng, Yao and Sadeh}]{feng2021aDesignSpace}
\bibinfo{author}{Feng, Y.}, \bibinfo{author}{Yao, Y.}, \bibinfo{author}{Sadeh,
  N.}, \bibinfo{year}{2021}.
\newblock \bibinfo{title}{A Design Space for Privacy Choices: Towards
  Meaningful Privacy Control in the Internet of Things}.
  \bibinfo{publisher}{Association for Computing Machinery},
  \bibinfo{address}{New York, NY, USA}.
\newblock \URLprefix \url{https://doi.org/10.1145/3411764.3445148}.
\bibitem[{Gerber et~al.(2018)Gerber, Gerber and
  Volkamer}]{gerber2018explaining}
\bibinfo{author}{Gerber, N.}, \bibinfo{author}{Gerber, P.},
  \bibinfo{author}{Volkamer, M.}, \bibinfo{year}{2018}.
\newblock \bibinfo{title}{Explaining the privacy paradox: A systematic review
  of literature investigating privacy attitude and behavior}.
\newblock \bibinfo{journal}{Computers \& Security} \bibinfo{volume}{77},
  \bibinfo{pages}{226--261}.
\newblock \URLprefix
  \url{https://www.sciencedirect.com/science/article/pii/S0167404818303031},
  \DOIprefix\doi{https://doi.org/10.1016/j.cose.2018.04.002}.
\bibitem[{Holtz et~al.(2011)Holtz, Nocun and Hansen}]{holtz2011towards}
\bibinfo{author}{Holtz, L.E.}, \bibinfo{author}{Nocun, K.},
  \bibinfo{author}{Hansen, M.}, \bibinfo{year}{2011}.
\newblock \bibinfo{title}{{Towards Displaying Privacy Information with Icons}},
  in: \bibinfo{booktitle}{Privacy and Identity Management for Life},
  \bibinfo{publisher}{Springer}. pp. \bibinfo{pages}{338--348}.
\bibitem[{ISO(2018)}]{ISO9241-11:2018}
\bibinfo{author}{ISO}, \bibinfo{year}{2018}.
\newblock \bibinfo{title}{{Ergonomics of human-system interaction -- Part 11:
  Usability: Definitions and concepts}}.
\newblock \bibinfo{type}{Standard} \bibinfo{number}{ISO 9241-11:2018}.
\bibitem[{Johansen and Fischer-H\"{u}bner(2019)}]{johansen2020theTR}
\bibinfo{author}{Johansen, J.}, \bibinfo{author}{Fischer-H\"{u}bner, S.},
  \bibinfo{year}{2019}.
\newblock \bibinfo{title}{{Making GDPR Usable: A Model to Support Usability
  Evaluations of Privacy}}.
\newblock \bibinfo{type}{Technical Report}. arXiv.
\newblock \URLprefix \url{arxiv.org/abs/1908.03503}.
\bibitem[{Johansen and Fischer-Hübner(2020)}]{johansen2020making}
\bibinfo{author}{Johansen, J.}, \bibinfo{author}{Fischer-Hübner, S.},
  \bibinfo{year}{2020}.
\newblock \bibinfo{title}{{Making GDPR Usable: A Model to Support Usability
  Evaluations of Privacy}}.
\newblock \bibinfo{journal}{IFIP Advances in Information and Communication
  Technology} \bibinfo{volume}{576}, \bibinfo{pages}{275–291}.
\newblock \DOIprefix\doi{10.1007/978-3-030-42504-3\_18}.
\bibitem[{Johansen et~al.(2020)Johansen, Pedersen, Fischer-H{\"u}bner,
  Johansen, Schneider, Roosendaal, Zwingelberg, Sivesind and
  Noll}]{johansen2020privacy}
\bibinfo{author}{Johansen, J.}, \bibinfo{author}{Pedersen, T.},
  \bibinfo{author}{Fischer-H{\"u}bner, S.}, \bibinfo{author}{Johansen, C.},
  \bibinfo{author}{Schneider, G.}, \bibinfo{author}{Roosendaal, A.},
  \bibinfo{author}{Zwingelberg, H.}, \bibinfo{author}{Sivesind, A.J.},
  \bibinfo{author}{Noll, J.}, \bibinfo{year}{2020}.
\newblock \bibinfo{title}{{Privacy Labelling and the Story of Princess Privacy
  and the Seven Helpers}}.
\newblock \bibinfo{type}{Technical Report}. arXiv.
\newblock \URLprefix \url{https://arxiv.org/abs/2005.08231}.
\bibitem[{Kvale(1994)}]{kvale1994interviews}
\bibinfo{author}{Kvale, S.}, \bibinfo{year}{1994}.
\newblock \bibinfo{title}{{Interviews: An introduction to qualitative research
  interviewing.}}
\newblock \bibinfo{publisher}{Sage Publications, Inc}.
\bibitem[{Mishler(1990)}]{mishler1990validation}
\bibinfo{author}{Mishler, E.G.}, \bibinfo{year}{1990}.
\newblock \bibinfo{title}{{Validation in Inquiry-Guided Research: The Role of
  Exemplars in Narrative Studies }}.
\newblock \bibinfo{journal}{Harvard Educational Review} \bibinfo{volume}{60},
  \bibinfo{pages}{415--443}.
\bibitem[{Patton(1999)}]{patton1999enhancing}
\bibinfo{author}{Patton, M.Q.}, \bibinfo{year}{1999}.
\newblock \bibinfo{title}{{Enhancing the Quality and Credibility of Qualitative
  Analysis}}.
\newblock \bibinfo{journal}{{Health Services Research}} \bibinfo{volume}{34},
  \bibinfo{pages}{1189--1208}.
\bibitem[{Preece et~al.(2015)Preece, Rogers and Sharp}]{preece2015interaction}
\bibinfo{author}{Preece, J.}, \bibinfo{author}{Rogers, Y.},
  \bibinfo{author}{Sharp, H.}, \bibinfo{year}{2015}.
\newblock \bibinfo{title}{{Interaction design: beyond human-computer
  interaction}}.
\newblock \bibinfo{publisher}{{John Wiley \& Sons}}.
\bibitem[{Spiekermann et~al.(2001)Spiekermann, Grossklags and
  Berendt}]{spiekermann2001privacy}
\bibinfo{author}{Spiekermann, S.}, \bibinfo{author}{Grossklags, J.},
  \bibinfo{author}{Berendt, B.}, \bibinfo{year}{2001}.
\newblock \bibinfo{title}{{E-privacy in 2nd Generation E-Commerce: Privacy
  Preferences versus actual Behavior}}, in: \bibinfo{booktitle}{Proceedings of
  the 3rd ACM conference on Electronic Commerce}, pp. \bibinfo{pages}{38--47}.
\bibitem[{Tesfay et~al.(2018)Tesfay, Hofmann, Nakamura, Kiyomoto and
  Serna}]{tesfay2018PrivacyGuide}
\bibinfo{author}{Tesfay, W.B.}, \bibinfo{author}{Hofmann, P.},
  \bibinfo{author}{Nakamura, T.}, \bibinfo{author}{Kiyomoto, S.},
  \bibinfo{author}{Serna, J.}, \bibinfo{year}{2018}.
\newblock \bibinfo{title}{Privacyguide: Towards an implementation of the eu
  gdpr on internet privacy policy evaluation}, in:
  \bibinfo{booktitle}{Proceedings of the Fourth ACM International Workshop on
  Security and Privacy Analytics}, \bibinfo{publisher}{Association for
  Computing Machinery}, \bibinfo{address}{New York, NY, USA}. p.
  \bibinfo{pages}{15–21}.
\newblock \URLprefix \url{https://doi.org/10.1145/3180445.3180447},
  \DOIprefix\doi{10.1145/3180445.3180447}.
\bibitem[{Whitten and Tygar(1999)}]{whitten1999johnny}
\bibinfo{author}{Whitten, A.}, \bibinfo{author}{Tygar, J.D.},
  \bibinfo{year}{1999}.
\newblock \bibinfo{title}{{Why Johnny Can't Encrypt: A Usability Evaluation of
  PGP 5.0.}}, in: \bibinfo{booktitle}{{USENIX Security Symposium}}.

\end{thebibliography}

\newpage
\appendix

\section{Demographic questions addressed to all participants, irrespective of group}\label{appendix_demographics_q}
\ \\

\hspace{-1.5cm}\includegraphics[width=21cm]{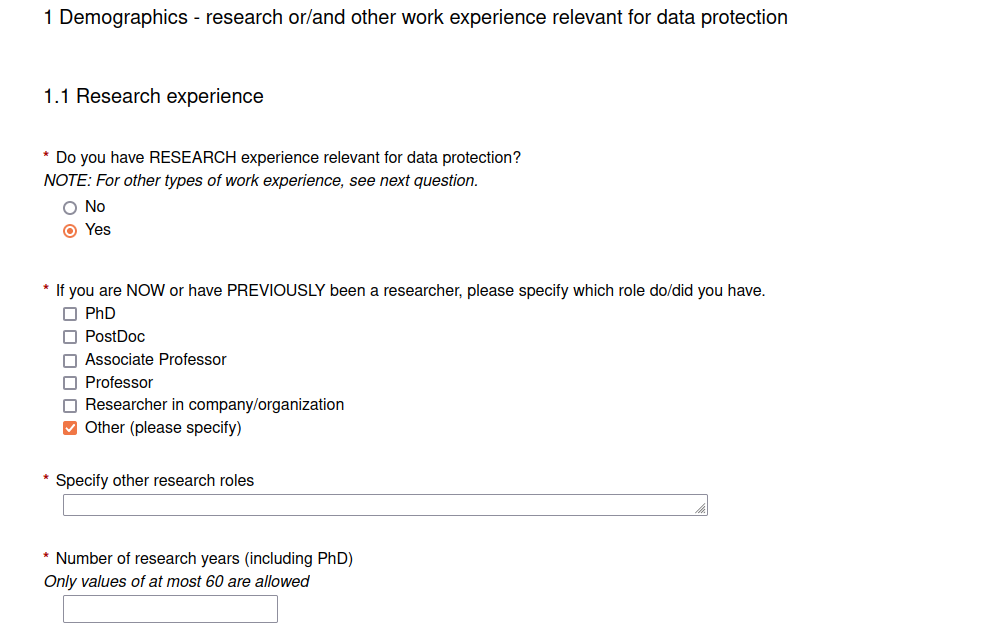}

\hspace{-1.5cm}\includegraphics[width=19cm]{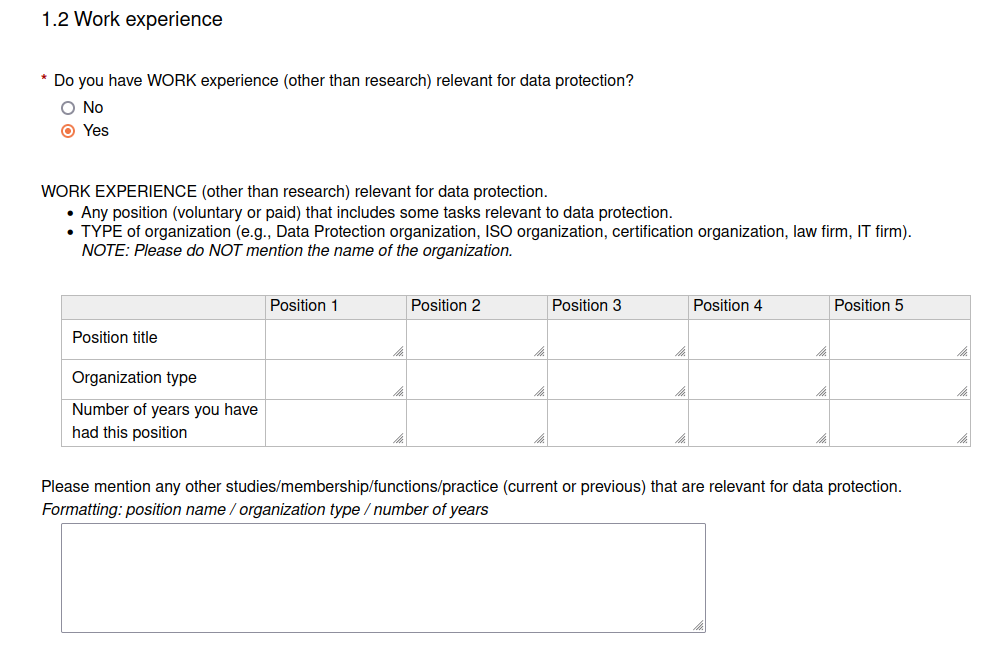}

\hspace{-1.5cm}\includegraphics[width=21cm]{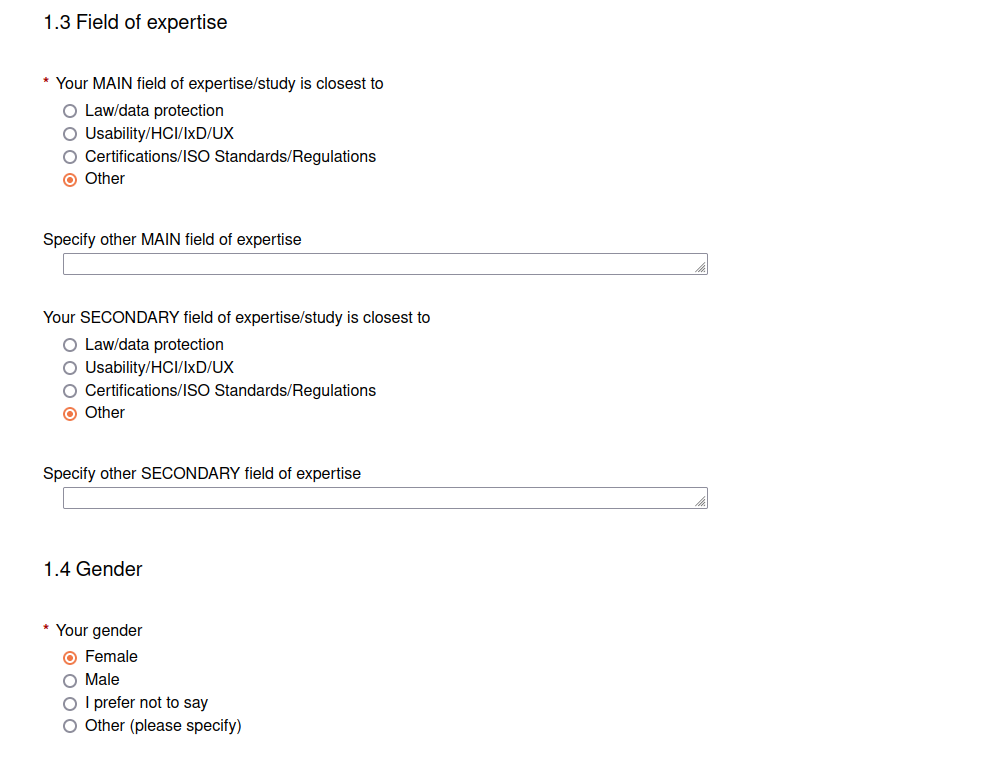}

\newpage
\section{Topics discussed with the participants in all three groups}\label{appendix_perspectiveExperienceOnUsability_q}
\hspace{-1.5cm}\includegraphics[width=22cm]{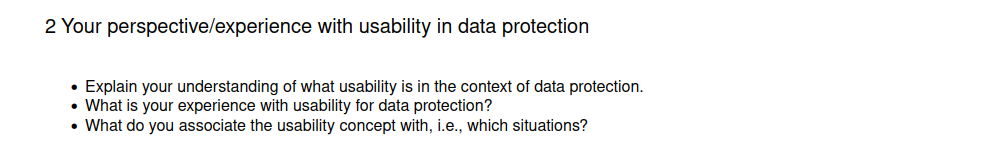}

\section{Topics discussed with the `usability group'}
\subsection{Usable Privacy Definition}\label{appendix_UPDefinition}
\hspace{-1.5cm}\includegraphics[width=21cm]{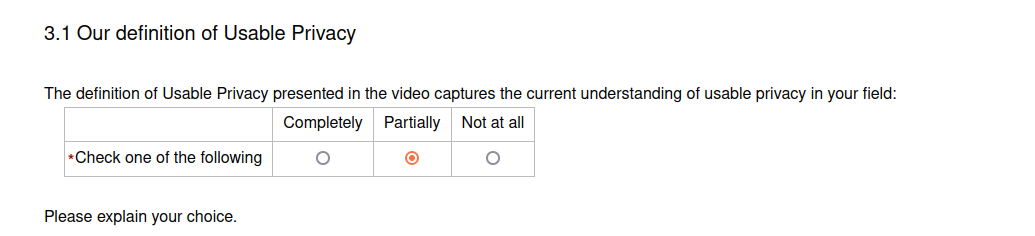}

\newpage
\subsection{Usable Privacy Criteria}\label{appendix_UPCriteria}
\ \\

\hspace{-1.5cm}\includegraphics[width=19cm]{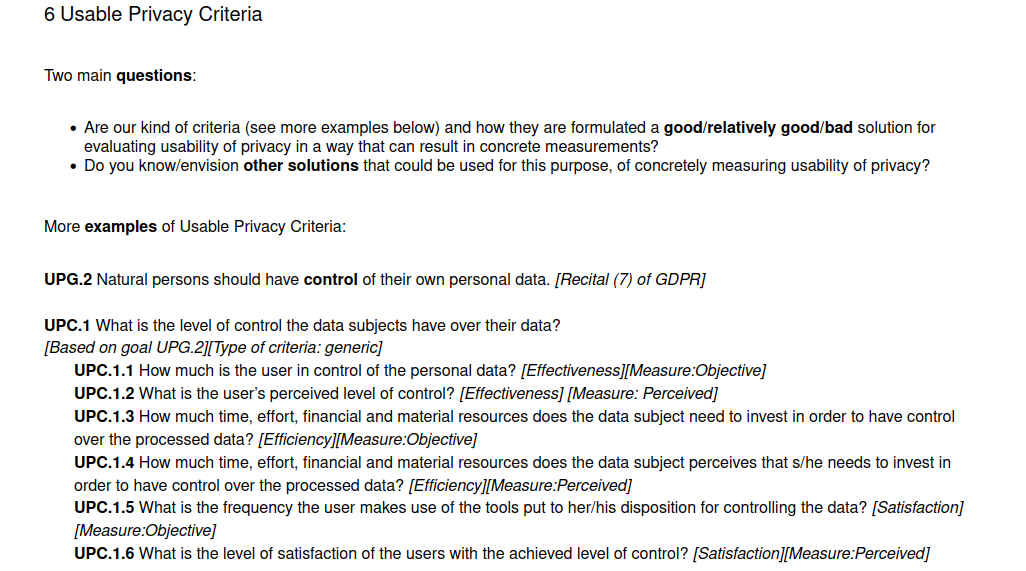}

\hspace{-1.5cm}\includegraphics[width=19cm]{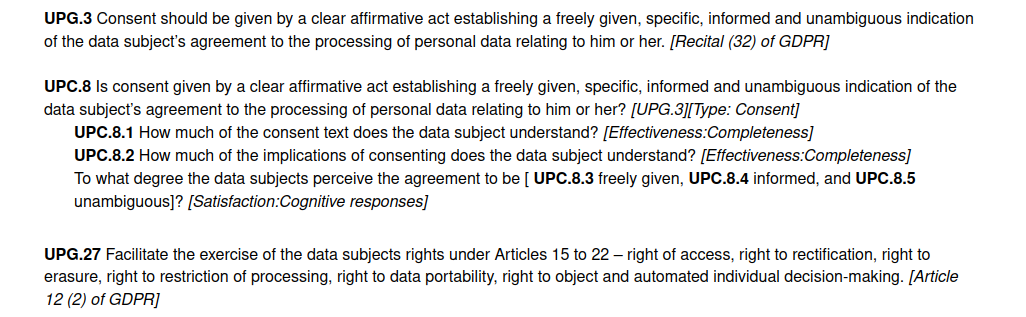}

\hspace{-1.5cm}\includegraphics[width=19cm]{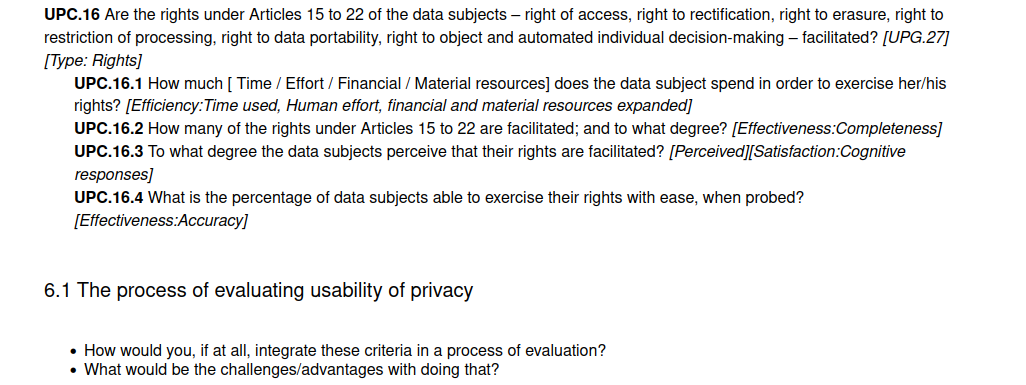}

\newpage
\section{Topics discussed with the `law group'}
\subsection{Usable Privacy Goals}\label{appendix_UPGoals}
\ \\

\hspace{-1.5cm}\includegraphics[width=19cm]{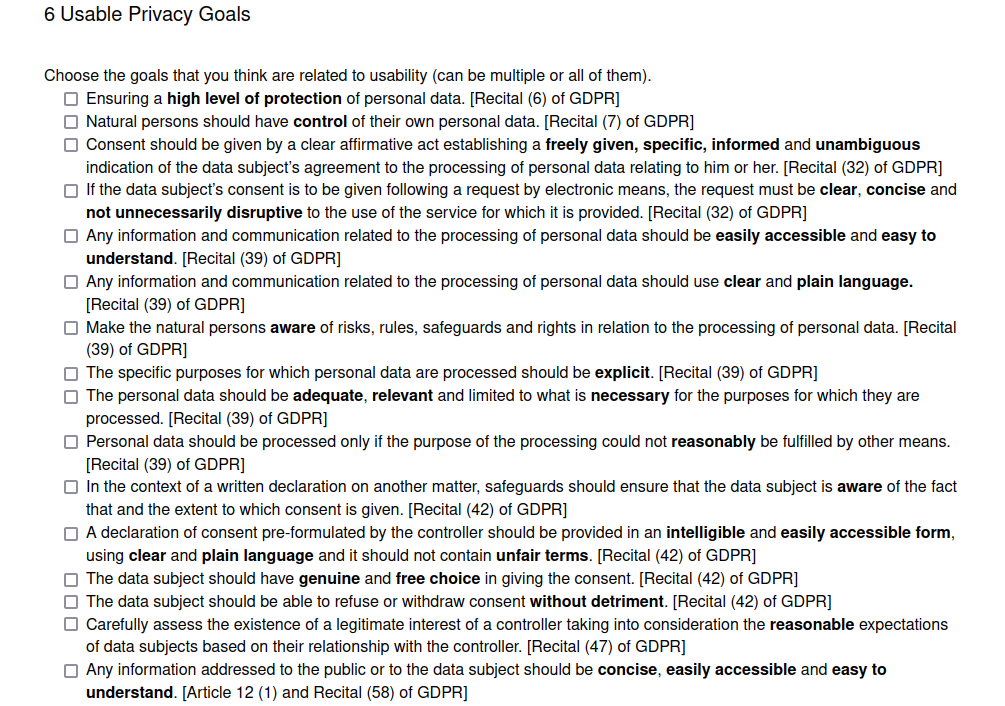}

\hspace{-1.5cm}\includegraphics[width=19cm]{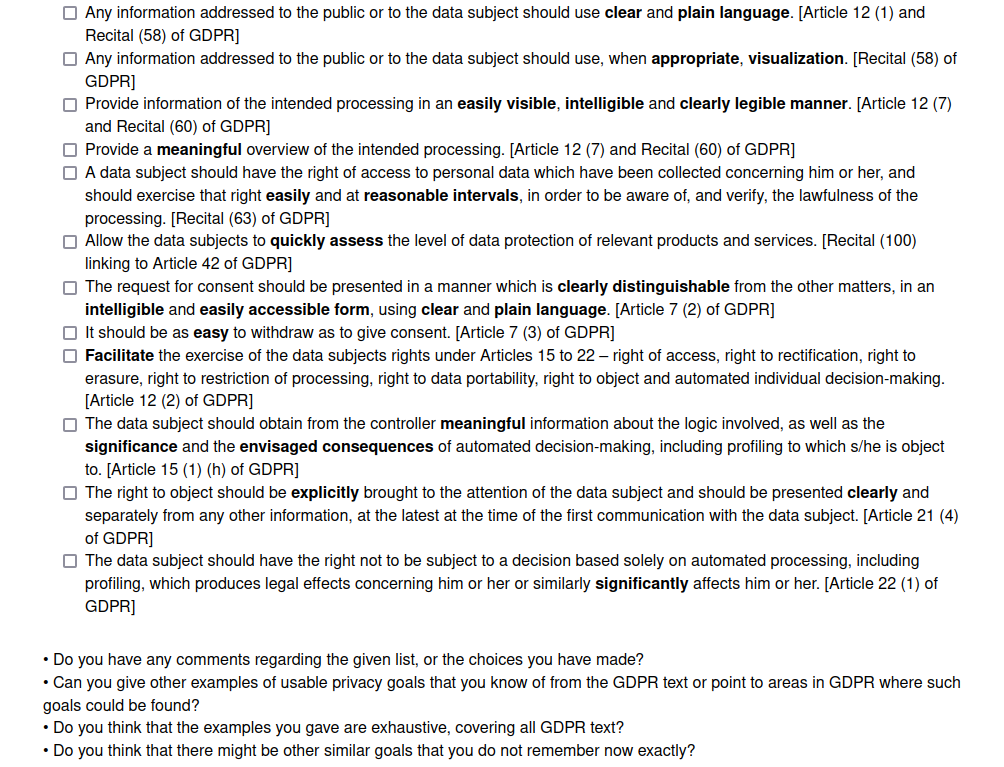}

\newpage
\section{Topics discussed with the `certifications group'}
\subsection{Evaluating and measuring usability of privacy}\label{appendix_evaluatingMeasuring}
\ \\

\hspace{-1.5cm}\includegraphics[width=19cm]{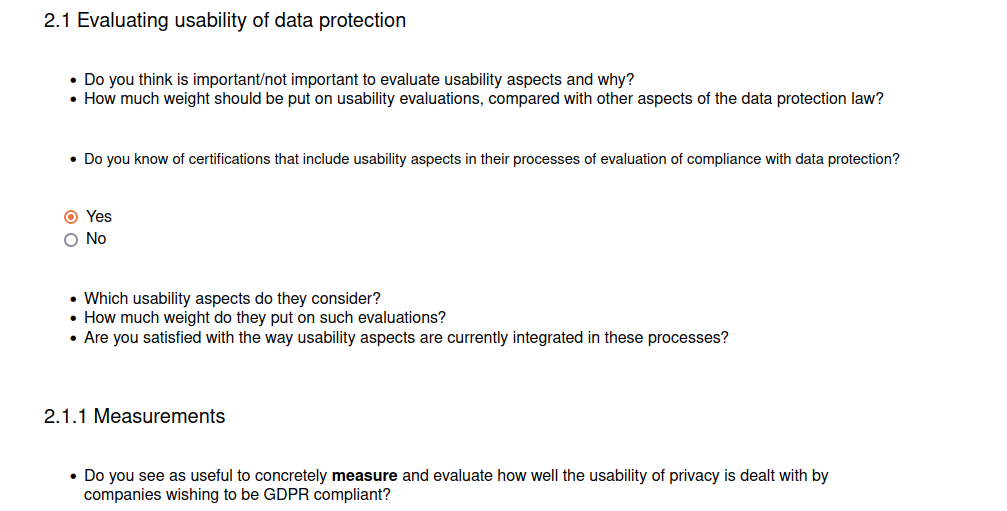}

\newpage
\subsection{Usable Privacy Cube}\label{appendix_UPCube}
\ \\

\hspace{-1.5cm}\includegraphics[width=19cm]{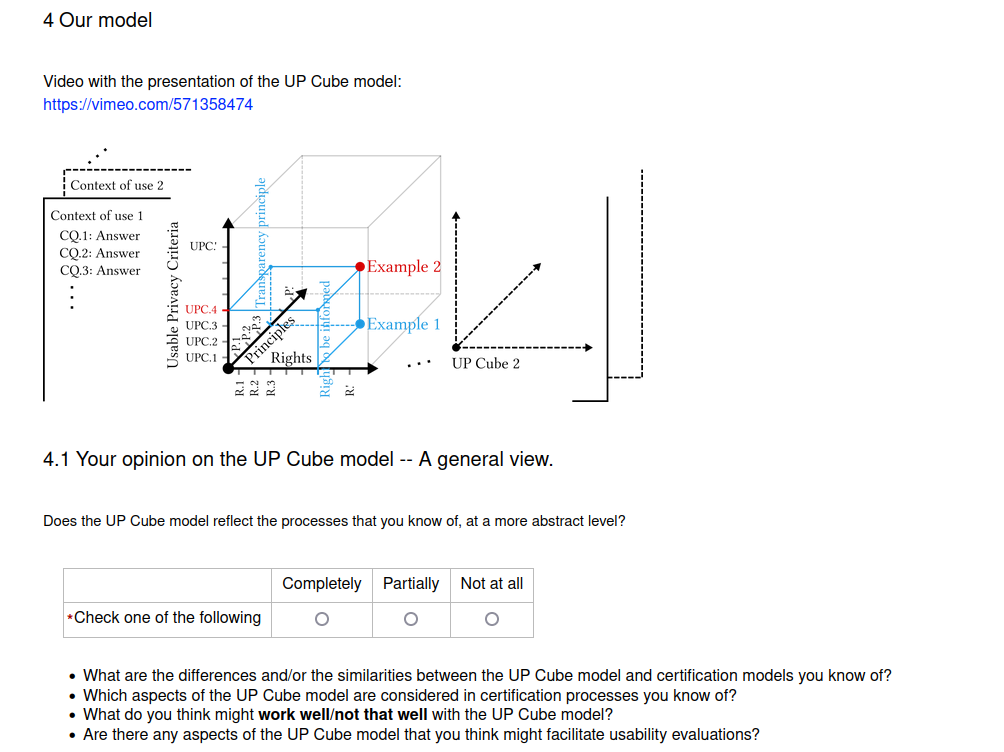}

\hspace{-1.5cm}\includegraphics[width=19cm]{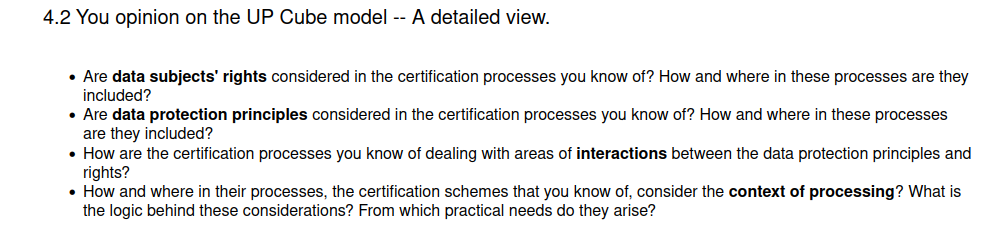}

\end{document}